\documentclass[aps,showpacs,twocolumn,floatfix,amsmath,preprintnumbers,nofootinbibe,secnumarabic]{revtex4} 

\usepackage{amsmath,amssymb,hyperref,url}

 \def\jrn#1#2#3#4#5#6{#1. #2. \textit{#3} \textbf{#6}, \textit{#4}, #5.} \def\andd{.; } \def\Ref{}   \def\eq{Equation } \def\eqs{Equations } \def\eqref#1{(\ref{#1})} \def\Sec{Section}  

\def\Ref{Ref.}  \def\eq{Eq. } \def\eqs{Eqs. }  

\def\scn#1#2{\section{#1}\lb{#2}} \def\sscn#1#2{\subsection{#1}\lb{#2}}

\def\bfl{\begin{flushleft}}
\def\efl{\end{flushleft}}
\def\bfr{\begin{flushright}}
\def\efr{\end{flushright}}
\def\bc{\begin{center}}
\def\ec{\end{center}}
\def\be{\begin{equation}}
\def\ee{\end{equation}}
\def\bse{\begin{subequations}}
\def\ese{\end{subequations}}
\def\ba{\begin{eqnarray}}
\def\ea{\end{eqnarray}}
\def\baa#1{\begin{array}{#1}}
\def\eaa{\end{array}}
\def\bw{\begin{widetext}}
\def\ew{\end{widetext}}
\def\nn{\nonumber }
\def\lb#1{\label{#1}}
\def\bit{\begin{itemize}}
\def\eit{\end{itemize}}
\def\bco{}
\def\bcs{\begin{cases}}
\def\ecs{\end{cases}}

\def\Der#1#2{\frac{d #1}{d #2}}

\def\Exp#1#2{\, \text{exp}^{#1}\left[#2 \right] }

\def\vena{\boldsymbol{\nabla}}
\def\vc#1{\mathbf{#1}}
\def\Tr{\text{Tr}}
\def\nndo{\hat{W}}
\def\ndo{\hat{\rho}}

\def\nnido{\nndo_\np}

\def\av#1{\langle #1 \rangle_W}
\def\avo#1{\langle #1 \rangle_\rho}
\def\avi#1{\langle #1 \rangle_{W_\np}}

\def\disop{\hat{{\cal D}}}

\def\dfn{\equiv}
\def\elp{{\phi}}
\def\np{\aleph}
\def\gyr{R_s}

\def\uyy{U_2}
\def\uy{U_1}
\def\huyy{\hat{U}_2}
\def\ez{{\bf e}_z}
\def\su{\mho}
\def\zfl#1{\overline{#1}} 

\def\avz#1{\av{#1}}
\def\Exp#1{\exp{\!(#1)}\,}

\def\pob#1{ \left\{ #1 \right\}_\text{c}}
\def\mob#1{ \boldsymbol{\{}\!\{ #1 \}\!\boldsymbol{\}}}
\def\mobs#1{ \boldsymbol{\{}\!\{\!\{ #1 \}\!\}\!\boldsymbol{\}}}

\def\ssss#1{\grave{#1}}
\def\ssw#1{\ssss{#1}}
\def\sss#1{\ssss{#1}}
\def\esds{\ssss{{\cal Z}}_\text{dw}}
\def\eszs{\ssss{{\cal Z}}_\text{zf}}
\def\ends{\ssss{{\cal E}}_\text{dw}}
\def\enzs{\ssss{{\cal E}}_\text{zf}}

\begin{document}




\preprint{\scriptsize \it
17th Russian Gravitational Conference - International Conference on Gravitation, Cosmology and Astrophysics (RUSGRAV-17)~}

\preprint{\small \footnotesize Universe \textbf{6}, 216 (2020)   
\ \ 
[\href{https://doi.org/10.3390/universe6110216}{DOI: 10.3390/universe6110216}]
}

\title{~\\
Density operator approach to turbulent flows in plasma and atmospheric fluids}


\author{Konstantin G. Zloshchastiev}
\email{https://orcid.org/0000-0002-9960-2874} 
\affiliation{Institute of Systems Science, Durban University of Technology, P.O. Box 1334, Durban 4000, South Africa}


\begin{abstract}
We formulate a statistical wave-mechanical approach to describe dissipation and instabilities in two-dimensional 
turbulent flows of magnetized plasmas and atmospheric fluids, such as drift and Rossby waves.
This is made possible by the existence of Hilbert space, associated with the electric potential of plasma or stream function of atmospheric fluid.
We therefore regard such turbulent flows as macroscopic wave-mechanical phenomena, driven by the non-Hermitian Hamiltonian operator we derive, whose anti-Hermitian component is attributed to an effect of the environment.
Introducing a wave-mechanical density operator for the statistical ensembles of waves, we formulate master equations
and define observables: such as the enstrophy and energy of both the waves and zonal flow as statistical averages.
We establish that our open system can generally follow two types of time evolution, depending on whether the environment hinders or assists the system's stability and integrity.
We also consider a phase-space formulation of the theory, including the geometrical-optic limit and beyond, and study the conservation laws of physical observables.
It is thus shown that the approach predicts various mechanisms of energy and enstrophy exchange between drift waves and zonal flow, which were hitherto overlooked in models based on wave kinetic equations. 
\end{abstract}


\date{received: 14 September 2020}



\maketitle

\scn{Introduction}{s:intr}
Zonal flow (ZF)
is a turbulence-related phenomenon
observed in physical systems
which appear to be very different,
such as
planetary atmospheres, protoplanetary discs, 
and astrophysical and laboratory plasmas; 
an extensive list of literature can be found in recent reviews \cite{pkf20,dii05}. 
For example,  Rossby waves (RW)
occur in planetary atmospheres \cite{hh94,vs05,wo09}, 
magneto-rotational turbulence occurs in accretion discs \cite{jyk09,kl13}, and drift-wave (DW) turbulence
is observed in fusion plasmas \cite{hbd93,ho99,fu09}. 
Yet another range of phenomena related to zonal flows are flow instabilities, 
jets and transitions to the chaos regime \cite{cs62,fi03,sy12,pk13,cfi14}.

The first attempts to describe  turbulence in plasma were made as early as the 60's, in works by Vedenov, Drummond and Pines, Kadomtsev, and others \cite{kabook}.  
About a decade later, Hasegawa and Mima proposed a new approach to low-frequency turbulence in 
nonuniform 
strongly magnetized plasma 
\cite{hm78}.  
In their approach, 
the fluid approximation and a perturbation approach were applied, resulting in the nonlinear Hasegawa-Mima equation
(HME), which has since become a very popular model in the
theory of zonal flows, drift waves and their interactions \cite{sd99,kk00,cnq14}. 

In a different branch of physics, the theory of atmospheric fluids, 
RW-related turbulence occurs as a result of fluid rotation,
one example being the
turbulence in Earth's atmosphere caused by the planet's rotation.
Its evolution equations turn out to be very similar to the
Hasegawa-Mima equation \cite{cs62}. 
In such systems, the role of drift waves is played by the Rossby waves,
which allows direct analogies between plasmas and atmospheric flows \cite{gd15,p16,rd16}.

It is known that turbulence transfers energy from large scales to smaller ones, thus causing it to dissipate. 
For instance, it is this kind of dissipation which leads to difficulties with magnetic fusion confinement.
Therefore,
it is important to study dissipative processes associated with drift or Rossby waves and zonal flows.
There are many approaches to describing dissipative (open) systems, 
each with different 
underlying assumptions and various
degree of rigor.

In our case, it is of great help that a sufficiently general DW/RW system can be mapped onto a wave-mechanical
system described by a Hamiltonian operator, \textit{akin} to a Schr\"odinger equation in quantum mechanics,
 the only difference being that this operator turns out to be non-Hermitian.
This non-Hermiticity reaffirms the fact that we are dealing with an open system \cite{fbook}.
By virtue of this flow-Schr\"odinger analogy, one can apply a  quantum-statistical technique based on master equations
with non-Hermitian Hamiltonians (NH), 
developed relatively 
recently \cite{sz13,sz14,sz14cor,ser15w,z15,sz15,sg16}.
This approach is proven to be robust for a large set of physical systems and phenomena, 
to mention just recent literature \cite{z16,z17adp,bk18,evg18,hh18,jg18,li18,wf18b,wf18a,se19,ev19,fs19,gc19,hrb19,kr19,lz19,da20,jvl20,bg20,cc20,aam20,cmm20,gb20,ghh20,lrw20,liu20,pb20,fvs20,wf20c,wf20q,wjs20,hgg20}. 
The advantage of this approach is that 
it makes the whole theory of dissipative DW/RW systems a subset of the modern theory of open systems;
which is statistical mechanical by nature, and describes both microscopical and macroscopic phenomena,
as well as interplays between them from basic quantum-mechanical principles \cite{bpbook}. 

The outline of the paper is as follows.
In Section \ref{s:mod}, we give a brief introduction to the generalized Hasegawa-Mima model,
and the underlying assumptions thereof. 
In Section \ref{s-ana}, we formulate a mapping between the HME flow equations and  wave
equations of a Schr\"odinger type,
which allows us to describe wave-mechanical phenomena in zonal flows. 
In Section \ref{s:do}, we generalize the wave-mechanical approach from state vectors 
to their statistical ensembles by introducing
density operator and master equations.
We find that two types of time evolution are possible, described by non-normalized and normalized
density operators, and describe differences between the two.
In Sections \ref{s:ww} and \ref{s:wws}, we formulate
a phase  space approach for finding solutions of master equations in, respectively,
non-normalized and normalized
density operator cases.
Section \ref{s:con} summarizes our findings and presents our conclusions.



\scn{Hasegawa-Mima model}{s:mod}
Let us derive a generalized Hasegawa-Mima equation as an example,
which is widely used to describe electrostatic two-dimensional turbulent 
flows \cite{sd99,kk00}.
These flows can occur in  magnetized
plasma, which exhibits the drift-wave  turbulence.
The plasma itself is usually assumed to be collisionless and quasi-neutral, its magnetic field 
$\vc B$ being uniform,
ion temperature being much smaller than the electron temperature,
and electrons following  Boltzmann distribution \cite{bibook,chbook}.

As mentioned in our introduction,
such flows can also occur in atmospheric
fluids on a rotating planet, where the role of drift waves is played
by Rossby waves. 
We are going to describe both types of physical systems using the same approach, the only difference being 
the values of some parameters,
therefore
our results should be applicable to both types of waves.

\sscn{Hasegawa-Mima equation for plasma}{s:moda}
We assume a conventional geophysical
coordinate system, where $\vc x = (x, y)$ are coordinates on a two-dimensional plane,
such that the $x$ axis lies in the zonal flow  direction, 
and the $y$-axis points in the
direction of the local gradient of the plasma density (in atmospheric fluids,
it would be the Coriolis force parameter).

Let us derive a generalized Hasegawa-Mima equation
in the case of plasma being entrapped
in the magnetic field, which is directed along $z$-axis:
\be
\mathbf{B} = B \, \ez
,
\ee 
where $B = B_0$ is a constant magnitude
and
$\ez$ is a unit vector normal to the plane.
Plasma is assumed to be confined to a slab, which is perpendicular to the magnetic field, so that perturbations of density and potential can propagate only in the $(x,y)$ plane. 
It is inhomogeneous, with density gradient pointing along $x$-axis. 
 
From now on, we assume the cold ions approximation $T_i \ll T_e$. 
Electron and ion components' temperatures are assumed to be constant, therefore 
the electron density is
$n_e=n_0 \exp{(e \tilde\phi / T_e)}$,
where
$\phi =  \phi_0 + \tilde{\phi}$ being electric potential,
$n_0$ is  unperturbed density and $T_e$ is the electron temperature. 
If $ e \tilde{\phi} \ll T_e$, 
then $n_e\approx n_0(1+ e\tilde{\phi}/T_e)$. 
Keeping perturbations of electron density up to a first order only, $n_e = n_0+\tilde{n}_e$, we obtain 
\be
\frac{\tilde{n}_e}{n_0}=\frac{e\tilde{\phi}}{T_e},
\label{1-10}
\ee
therefore, in the formula $\phi = \phi_0 + \tilde{\phi}$ the first term vanishes.

In a slab plasma with $v_z \ll v_\perp$, with the electrostatic field being $\mathbf E =- \vena \phi$, 
where $\vena= \mathbf e_x \partial_x 
+ \mathbf e_y \partial_y$, the equation of motion  for an ion is
\be
m_i\frac{d}{dt}\mathbf v_{\perp}
=
q \left (-\vena \phi+\mathbf v_{\perp}\times \mathbf B 
\right)
,
\label{1-20}
\ee
where $m_i$ and $q$ are ion's mass and charge, respectively.
Applying a perturbation theory, 
\ba
\mathbf {v_{\perp}}
&=&
\mathbf {v_{\perp}}^{(0)}+\lambda \mathbf {v_{\perp}}^{(1)}+\lambda^2 \mathbf {v_{\perp}}^{(2)}+ ...
,\nn\\
\phi 
&=& \phi_0(r)+\lambda \tilde{\phi} (r,t)+ ... 
,\nn\ea
we can write \eq \eqref{1-20} as 
\bw
\ba
&&
\frac{1}{\omega_{ci}}\frac{d}{dt}
\left(
\mathbf {v_{\perp}}^{(0)}+\lambda \mathbf {v_{\perp}}^{(1)}+\lambda^2 \mathbf {v_{\perp}}^{(2)}+ ... 
\right)
=
\frac{1}{B}
\left[
-\vena 
\left(
\lambda \tilde{\phi} (r,t)+... 
\right)
+
\left(
\mathbf {v_{\perp}}^{(0)}+\lambda \mathbf {v_{\perp}}^{(1)}+\lambda^2 \mathbf {v_{\perp}}^{(2)}+ ... 
\right)
\times \mathbf B
\right]\!,~~~
\label{1-30}
\ea
\ew
where $\omega_{ci} = q B/m_i$ is the ion's cyclotron frequency.
To a zeroth order of perturbation theory, we thus obtain the equation:
\be
\frac{1}{\omega_{ci}}\frac{d \mathbf v_{\perp}^{(0)}}{dt}
=
\frac{1}{B} \left (\mathbf v_{\perp}^{(0)}\times \mathbf B \right),
\label{1-40}
\ee
whose solution describes rotary motion, so that we have
$ 
\mathbf v_{\perp}^{(0)}=0.
$ 

In the first order, one can write
\be
\frac{1}{\omega_{ci}}\frac{d \mathbf v_{\perp}^{(1)}}{dt}=
-\frac{1}{B} \left (\vena \tilde \phi-\mathbf v_{\perp}^{(1)}\times \mathbf B \right),
\label{1-60}
\ee
whose solution is given by
 \be
\mathbf v_{\perp}^{(1)}= 
-\frac{1}{B}\vena \tilde{\phi} \times \ez - \frac{1}{\omega_{ci}B}\frac{D}{D t}\vena\tilde \phi
,
\label{1-70}
\ee
where 
we 
denoted the convective derivative as
\[
\frac{D}{Dt}
\dfn
\partial_t
-\frac{1}{B}(\vena \tilde \phi \times \ez) \cdot \vena
,\]
where by $\partial_t$ we denote a partial time derivative. 
Ignoring higher-order corrections, 
we obtain
\ba
\mathbf v_{\perp} =
\mathbf v_{\perp}^{(1)}&=&
-\frac{1}{B}\vena \tilde \phi \times \ez 
\nn\\&&
-
\left[
\partial_t
-
\frac{1}{\omega_{ci} B^2}
(\vena \tilde \phi \times \ez) \cdot \vena \right] \vena \tilde \phi
.~~
\label{e:vappr}
\ea
By substituting $n_i=n_0+\tilde{n}_i$  into the continuity equation for the ion component's fluid,
\be
\frac{\partial n_i}{\partial t}+\vena \cdot \left (n_i \mathbf v_{\perp} \right)=0
,
\label{1-90}
\ee
and using the quasi-neutrality condition $\tilde{n}_i=\tilde{n}_e$,
we obtain the following equation 
\be
\partial_t
\left(\frac{\tilde{n}_e}{n_0}\right)
+\mathbf v_{\perp} \cdot 
\left (
\frac{\vena n_0}{n_0}+\frac{\vena \tilde{n}_e}{n_0}
\right ) 
+\left (1+
\frac{\tilde {n}_e}{n_0}
\right)
\vena \cdot \mathbf{v}_\perp =0
,
\label{1-100}
\ee
which transforms, using \eqs \eqref{1-10} and \eqref{e:vappr},
and neglecting terms which are higher than the first order, 
into the equation
\bw
\be 
\partial_t\frac{e \tilde{\phi}}{T_e}
-\frac{1}{B}
\left(
\vena \tilde{\phi}\times \ez
\right)
\cdot 
\left(
\frac{\vena n_0}{n_0}+\frac{e \vena \tilde{\phi}}{T_e}
\right ) 
-
\left(
1+\frac{e \tilde{\phi}}{T_e}
\right)
\vena \cdot 
\left \{
\frac{1}{B} \left (\vena \tilde{\phi}\times \ez\right)  
+
\frac{1}{\omega_{ci} B}
\left [
\partial_t-\frac{1}{B}
\left(
\vena \tilde{\phi}\times \ez
\right)
\cdot \vena 
\right] 
\vena \tilde{\phi}
\right \}
=0.
\label{1-110}
\ee
\ew
In this equation,  
the term $(\vena \tilde{\phi} \times \ez) \cdot \vena \tilde{\phi}$ 
vanishes due to orthogonality,
while the terms with
$(e \tilde \phi /T_e) \vena \cdot$ can be neglected as being higher-order small. 
Multiplying \eq \eqref{1-90} by $- T_e/e$ and recalling an expression for the ion's cyclotron (gyro)radius, 
\be
\gyr
=
\frac{c_s}{\omega_{ci}}
=
\frac{1}{\omega_{ci}}
\sqrt{
\frac{T_e}{m_i}
}
, 
\ee
where $c_s = T_e/m_i$ being an ion's 
characteristic
sound speed, 
we obtain
\ba
&&
 \partial_t
 \left[
 \left(
 \rho^2_s \vena^2-1
 \right)
 \tilde \phi
 \right] 
 + 
 \left( 
 \ez \times \frac{\vena \tilde \phi}{B}
 \right) 
 \cdot \vena 
 \Big[ 
 \big (
 \rho^2_s \vena^2-1
 \big )
 \tilde \phi
 \Big]
 \nn\\&&\qquad\qquad\qquad\qquad
-
 \frac{T_e}{|e| B}\frac{\partial \ln n_0}{\partial y}\frac{\partial \tilde \phi}{\partial x}
=
0
.~~~~
 \label{1-120}
\ea
In terms of dimensionless values
$  \acute{t} = \omega_{ci} t$,  $( \acute{x},  \acute{y}) = (x/ \gyr, y / \gyr) $ and $ \acute{\phi} = e \tilde \phi / T_e$,
we thus obtain 
\[ 
 \frac{\partial}{\partial \acute{t}} \Big [ (\vena^2-1) \acute{\phi}\Big ] +(\ez \times \vena \acute{\phi}) \cdot \vena \big [ (\vena^2 - 1) ]\acute{\phi}\big] - \frac{\partial \ln n_0}{\partial \acute{x}} \frac{\partial \acute{\phi}}{\partial \acute{x}}
=
0
,
\] 
therefore
\be
\frac{\partial  w }{\partial \acute {t}}
+
\mathbf v
\cdot \vena  w 
+
\beta \frac{\partial \phi ^\prime}{\partial x^\prime}
=
Q,
\label{1-140}
\ee
 where 
$Q (\vc x, t)$ is an added external forces' and dissipation term, 
$
 \beta \dfn V_*/c_s$ 
is a positive constant parameter,
$ 
 V_* \dfn
 - (T_e/| e| B)
 \,
 \partial_ {\acute{x}}
 \ln n_0  $,
 and 
 $ 
 w = (\vena ^{2}-1)\acute{\phi}
 $ 
is a vorticity
defined as a measure of local angular velocity,
see Appendix \ref{s:appw} for details.

Placing both plasma and atmospheric flow cases on the same footing, 
one can write the generalized Hasegawa-Mima equation
in the final form:
\be\lb{e:ghme}
\partial_t w +
\vc v \cdot \vena w + \beta \partial_x \elp 
= Q
,
\ee
where
the generalized vorticity $w (\vc x, t)$
can be
written in the form:
\be
w
=
\left(
\vena^2 - L_D^{-2} \hat\alpha
\right)
\elp 
, \ee
where 
$\vc v = \ez \times \vena \elp$ is the fluid velocity on the $(x, y)$ plane,
$\hat\alpha$ is a preselected operator whose value equals  
the unit operator $\hat I$ (DW case) or to zero (RW case),
$\elp (\vc x, t) $ is the electric potential or stream function,
$L_D$ is the plasma sound radius or deformation radius.
Here, and below,
we work with the dimensionless values, but omit primes for brevity.

\sscn{Fluctuations}{s:modf}
Let us define zonal averages according to the formula 
$\zfl{A}(y)
\dfn
 L_x^{-1} \int_{0}^{L_x} A \, dx $, 
where $L_x$ is the system's extent along $x$ axis.
From now on
we assume that a tilde and bar refer to, respectively, fluctuation and zonal-averaged values.
We therefore start with the following expansion \cite{rd16}:
\ba
w &=& \tilde{w}(x,y,t) + \zfl{w}(y,t),
\label{eq:vortComp}
 \\
\phi&=& \tilde \phi(x,y,t) 
+
\zfl{\phi}(y,t),
\label{eq:potComp}
\ea
where 
\be
\tilde{w}
=
(\vena^2 - 1)\tilde{\phi}
,
\quad
\zfl{w}
=
\vena^2 \zfl{\elp}
=
\partial^2_y \zfl{\elp}
.
\label{eq:wbar}
\ee
Note that if we want to restore the parameter $L_D$, 
we would obtain
\be
\tilde{w}
=
(\vena^2 - L_D^{-2})\tilde{\elp}
, \quad
\zfl{w}
=
\vena^2 \zfl{\elp}
= \partial^2_y \zfl{\elp}
.
\label{e:wbar}
\ee
Furthermore, \eq \eqref{1-140} can be written 
in the
quasilinear approximation
as
\ba
 &&
\partial_t \tilde{w}
+
\tilde{\mathbf v} \cdot \vena \zfl{w}
+
\zfl{\mathbf v} \cdot \vena \tilde{w}
+
\beta \partial_x \tilde{\phi}
=
\tilde{Q},
\label{eq:fluc}
\\&&
\partial_t \zfl{w}
+
\overline {\tilde{\mathbf v}\cdot \vena \tilde w}
=
\zfl{Q}
,
\label{eq:ZAver}
\ea
because eddy-eddy interactions can be omitted \cite{sy12}.
Fluid velocity can be written in terms of its components as 
$\mathbf v = - \mathbf e_x \partial_y \phi + \mathbf e_y \partial_x \phi$,
therefore,
its fluctuating and zonal-averaged components are, respectively: 
\be
  \tilde{\mathbf v}
=
\mathbf{e}_x \tilde{v}_x+ \mathbf{e}_y \tilde{v}_y
, \quad
\zfl{\mathbf{v}}
=
 \mathbf{e}_x U,
 \label{eq:vtilde}
 \ee
 where  $\tilde{v}_x =-\partial_y \tilde{\phi}$,  $\tilde{v}_y =\partial_x \tilde{\phi}$, and 
\be\lb{e:udef}
U (y,t) = -\partial_y \zfl\elp
\ee 
is the $x$th component of a ZF velocity $\zfl{\vc v} = \vc{e}_x U$.

With the use of \eqs \eqref{eq:wbar}, \eqref{eq:vtilde}
and due to independence of $\zfl{\elp}$ from $x$,
the second term of \eq \eqref{eq:fluc} can be rewritten as	
 \be  
 \tilde{\mathbf v} \cdot \vena \zfl{w}
 =
 -
 \partial_y \tilde{\elp} \partial_x \partial^2_y
 \zfl{\elp}
 +
\partial_x \tilde{\elp} \partial^2_y \zfl{\elp}
 =
 -
 \partial_x \tilde{\elp} \uyy,
 \label{eq:secondinfluc}
\ee 
where we used the notation $U_k \dfn \partial^k U/\partial y^k = - \partial^{k+1} \, \zfl{\elp}/\partial y^{k+1}$.
A third term in \eq \eqref{eq:fluc} can be written as
\be 
\zfl{\mathbf v} \cdot \vena \tilde{w}
=
U \partial_x \tilde{w}
.
\label{eq:thirdinfluc}
\ee 
Furthermore,
a second term in \eqref{eq:ZAver} yields
\[
\tilde{\mathbf v}\cdot \vena \tilde w
=
-\partial_y \tilde{\phi}\partial_x
\left(
\partial^2_x + \partial^2_y -1 
\right)
\tilde{\phi}
+
\partial_x \tilde{\phi}\partial_y
\left(
\partial^2_x + \partial^2_y -1 
\right)
\tilde{\phi}
,
\]
therefore
\ba
\overline {\tilde{\mathbf v}\cdot \vena \tilde w}
&=&
\frac{1}{L_x} \int_{0}^{L_x}-\partial_y \tilde{\phi}\,
\partial_x\!
\left(
\partial^2_x + \partial^2_y -1 
\right)
\tilde{\phi}\; dx
\nn\\&&+
\frac{1}{L_x} \int_{0}^{L_x} \partial_x \tilde{\phi}\,
\partial_y\!
\left(
\partial^2_x + \partial^2_y -1 
\right)
\tilde{\phi}\, dx
\nn \\
&=&
-
\frac{1}{L_x}\partial_y \int_{0}^{L_x}
\left(
\partial_x \tilde{\phi}\partial^2_x \tilde{\phi}
+
\partial_x \tilde{\phi}\partial^2_y \tilde{\phi}
-
\tilde{\phi} \partial_x \tilde{\phi}
\right)
 dx
\nn\\&=&
-\partial^2_y\, \overline{\tilde{v}_x\tilde{v}_y}
,
\label{eq:secondinZAver2}
\ea 
where we used integration by parts in the last line.
Substituting \eqs \eqref{eq:secondinfluc}, \eqref{eq:thirdinfluc} and  \eqref{eq:secondinZAver2} 
into \eqs \eqref{eq:fluc} and \eqref{eq:ZAver}, 
we obtain: 
\ba 
&&
\partial_t \tilde{w}
+
U \partial_x \tilde{w}
+
\left(
\beta -
\uyy
\right)
\partial_x \tilde{\elp}
=
\tilde{\zeta}
-
\mu_\text{dw}^{(0)} \tilde{w},
\label{eq:zRu6a}
\\&&
\partial_t U
+
\mu_\text{zf}^{(0)} U
=
-
\partial_y \overline{\tilde{v}_x\tilde{v}_y}
,
\label{eq:zRu6b}
\ea
where
$\tilde{\zeta}$ is an
external dissipation source with zero zonal average,
$\mu_\text{dw}^{(0)} $ and $\mu_\text{zf}^{(0)} $
are the constant parameters of the terms describing the simplest kinds of dissipation
of, respectively, driftons and zonal flows, caused by the environment. 

Using \eq \eqref{eq:wbar},
these equations can be also rewritten in terms of the function $\tilde{\elp}$:
\ba 
&&
(\vena^2 - 1)
\partial_t \tilde{\elp}
=
-
\left[
\beta -
\uyy
+
U (\vena^2 - 1)
\right]
\partial_x \tilde{\elp}
\nn\\&&\qquad\qquad\qquad\ \ \,
+
(\vena^2 - 1)
\left(
\tilde{\xi}
-
\mu_\text{dw}^{(0)} \tilde{\elp}
\right)
,
\label{eq:Ru6a}
\\&&
\partial_t U
+
\mu_\text{zf}^{(0)} U
=
\partial_y\! 
\left(
\overline{
\partial_y \tilde{\elp} \, \partial_x \tilde{\elp}
}
\right)
,
\label{eq:Ru6b}
\ea
where $\tilde{\xi}$ is an
external dissipative potential with zero zonal average \cite{fi03}.

\sscn{Observables}{s:modo}
The important observables of the model are enstrophy and energy,
which can be defined as the following integrals:
 \ba 
  \mathcal{Z}_\text{dw} 
  &\dfn& \frac{1}{2} \int d^2x\, \tilde w^2   
	= 
	\frac{1}{2} \int d^2x  \left[(\vena^2 - 1) \tilde{\elp} \right]^2  
 ,
 \label{e:ensdw}\\
  \mathcal{Z}_\text{zf}
  &\dfn&
  \frac{1}{2} \int d y \, \zfl{w}^2 
	 =
 \frac{1}{2}\int d y \, \uy^2
 ,\label{e:enszf}\\
  \mathcal{E}_\text{dw}
  &\dfn&
  - \frac{1}{2}\int d^2x \, \tilde{w} \, \tilde{\elp} 
	=
	\frac{1}{2}\int d^2x \, \tilde{\elp}\, (1- \vena^2) \tilde{\elp}
 ,\label{e:endw}\\
 \mathcal{E}_\text{zf}
 &\dfn&
 - \frac{1}{2} \int  d y \, \zfl{w}\, \zfl{\phi} 
 =
 \frac{1}{2}\int d y \, U^2 
 ,
 \label{e:enzf}
 \ea 
where 
subscripts indicate a corresponding component, either a drift-wave or a zonal-flow one.
The total enstrophy and energy are, respectively
 \be\lb{e:eetot}
\mathcal{Z}_\text{tot} 
 =   \mathcal{Z}_\text{dw} +
 \mathcal{Z}_\text{zf}
, \ \
\mathcal{E}_\text{tot} 
 =   \mathcal{E}_\text{dw} +
 \mathcal{E}_\text{zf}
 ,
\ee
which are usually expected to be conserved values in those physical systems,
and
which allow description in terms of wave kinetic equations (WKE).

\scn{Wave-mechanical analogy in zonal flows}{s-ana}
Building on the ideas presented in works \cite{wo09,p16,rd16},
let us formulate a formal mapping of \eq \eqref{eq:Ru6a} to 
an effective Schr\"odinger-like equation which is somewhat
analogous to a quantum-mechanical description.
For reasons, which will be later specified, around \eq \eqref{e:hbareff},
we shall refer to this analogy as \textit{wave-mechanical}.

Judging by the form of \eq \eqref{eq:Ru6a}, one can expect that its solutions
form the Hilbert space of normalizable smooth functions, whose inner product is defined as
an integral over the whole $(x,y)$ plane, $\int d^2 x$.
Postponing discussion of the physical implications of this until the end
of this section,
we introduce, in Dirac's bra-ket notations,   
a state vector $|\tilde{\elp}\rangle$,
such that
in the coordinate representation one obtains
$\tilde{\elp}(\vc{x}, t)
=  \langle \vc{x}|\tilde{\elp}\rangle$, similarly for $\tilde{\xi} (\vc{x}, t)$.

Using the resolution of identity,
$
\int d^2 x  | \vc{x} \rangle \langle \vc{x}| = \hat I
$,
we obtain
\be\lb{e:wnorm}
\langle \tilde{\elp} | \tilde{\elp}\rangle
\dfn
\np
=
\int \langle \tilde{\elp}| \vc{x} \rangle \langle \vc{x}|\tilde{\elp}\rangle d^2 x
=
\int \tilde{\elp}^2 d^2 x
,
\ee
where $\np (t)$ is a norm of a state vector $|\tilde{\elp}\rangle$. 
This state vector 
is almost what we are looking for, but the proper Schr\"odinger analogy requires 
a \textit{normalized} state vector (ray), which spans a projective Hilbert space.
We thus introduce
\be\lb{e:wtopsi}
| \Psi \rangle 
\dfn
\np^{-1/2}
| \tilde{\elp}\rangle
,
\ee
so that
$ 
\langle \Psi | \Psi \rangle
=
1
$. 

Furthermore,
using an operator of spatial translations in the $(x,y)$ plane, 
$ 
 \hat{\vc{p}}
 =-i\vena
=-i(\vc{e}_x \hat{p}_x+\vc e_y \hat{p}_y)
$, 
we can define the Hermitian operators 
$\hat{p}^2 = \hat{\textbf{p}} \cdot \hat{\textbf{p}} = - \vena^2$,
$\hat{p}^2_D = \hat{p}^2+L_D^{-2} \Leftrightarrow 1 - \vena^2 $
and $\hat{U}
\dfn U(\hat{y},t) $.
In terms of these operators, \eq \eqref{eq:Ru6a} 
can be rewritten in the Schr\"odinger-like form:
 \be 
 i \partial_t 
\langle \textbf{x}|\Psi \rangle 
= 
\hat{H} \langle \textbf{x}| \Psi \rangle + 
i \np^{-1/2} \langle \textbf{x}| \tilde {\xi} \rangle
,
 \label{eq:FlucQuantx}
 \ee 
or, in a general basis
 \be 
 i \partial_t 
|\Psi \rangle 
= 
\hat{H} | \Psi \rangle + 
i \np^{-1/2} | \tilde {\xi} \rangle
,
 \label{eq:FlucQuant}
 \ee 
 where the Hamiltonian operator is given by
 \ba
 \hat{H}
 &=&
\left(
\hat\su
+
\hat{p}_D^{-2} 
\huyy
 - \beta \hat{p}_D^{-2} 
\right)
\hat{p}_x
 -
 i 
\mu_\text{dw} \hat I
,\label{e:hamtot}
 \ea 
where  
we 
introduced a similarity transform of $\hat U$:
\ba
\hat\su
=
\hat{p}_D^{-2}
\hat{U}
\hat{p}_D^{2}
, \ea
used an identity
$[\hat{p}_D^{-2}, \hat{\textbf{p}}] 
=
-\hat{p}_D^{-2}
[\hat{p}_D^{2}, \hat{\textbf{p}}] 
\hat{p}_D^{-2}
= 0$,
and  denoted
\be\lb{e:mudw}
\mu_\text{dw} 
= 
\mu_\text{dw}^{(0)}
+
\frac{1}{2}
\frac{\dot \np}{\np}
,
\ee
where dot means an ordinary derivative with respect to time.

It should be noticed that the term 
$i \dot \np/(2 \np)$ occurs in the Hamiltonian,
caused by the transition 
from $| \tilde{w} \rangle$
to the proper state vector \eqref{e:wtopsi},
which indicates that dissipation exists even if 
$\mu_\text{dw}^{(0)}$ is zero.
In general, this term is time-dependent, except in two cases: 
when $\np$ is constant (in which case the term vanishes) or
when $\np$ depends on time exponentially (then the term becomes a constant
representing a rate of loss or gain).
Note that because the canonical Schr\"odinger picture \textit{per se} presumes no explicit time dependence of
a
Hamiltonian operator,
time evolution of our system at
the general function $\np (t)$ 
must be described, not by
a Schr\"odinger equation, but by a master equation
as will be discussed later, in Section \ref{s:dome}.

Note also, that in 
\Ref \cite{rd16}
the flow-Schr\"odinger analogy was proposed,
but using wavefunction 
$\tilde \omega (\vc{x}, t)$ instead.
From \eq \eqref{eq:wbar}, it is clear that $\tilde \omega$ 
is a surjective function of $\tilde\elp$
(due to the partial differentiation thereof),
which
should rather be interpreted as a source density.
All this suggests that the Hilbert space spanned by state vectors $ | \tilde\elp \rangle $,
or by projective rays $ | \Psi \rangle $,  
is the underlying one.
Therefore, it is not surprising that by using 
the $\tilde\elp$-associated Hilbert space, one 
obtains more general Hamiltonian, 
as will be demonstrated below,
and also
decreases
the number of inverse differential operators 
in the resulting evolution equations, both
for the state vector and function $U$.

Furthermore,
in terms of the normalized state vector,
some of the integral values of Section \ref{s:modo}
can be written in a more convenient form to use in the next section.
We thus obtain in Dirac's notations
\ba 
\mathcal{Z}_\text{dw} 
  &=& 
\frac{1}{2} \np \langle \Psi | \hat{p}^{4}_D  | \Psi \rangle
 ,
 \label{e:ensdwb}\\
  \mathcal{Z}_\text{zf}
  &=&
 \frac{1}{2}\int d y \, \uy^2
 ,\label{e:enszfb}\\
  \mathcal{E}_\text{dw}
  &=&
\frac{1}{2} \np 
\langle \Psi | \hat{p}^{2}_D  | \Psi \rangle
 ,\label{e:endwb}\\
 \mathcal{E}_\text{zf}
 &=&
 \frac{1}{2}\int d y \, U^2 
 ,
 \label{e:enzfb}
 \ea  
where we used \eq \eqref{e:wtopsi}. 
 
Furthermore,
one can see that the Hermitian adjoint of an operator \eqref{e:hamtot},
 \be
 \hat H^\dag
 =
\left(
\hat\su^\dag
+
\huyy
\hat{p}_D^{-2} 
 - \beta \hat{p}_D^{-2} 
\right)
\hat{p}_x
 +
 i 
\mu_\text{dw} \hat I
,
 \ee
does not coincide with \eq \eqref{e:hamtot}.
Therefore, our model belongs to a class of theories with non-Hermitian Hamiltonians,
usually referred as NH theories.

Furthermore, operator \eqref{e:hamtot}
can be decomposed into its Hermitian and anti-Hermitian parts: 
\be
\hat H =
\hat H_+ 
+  \hat H_-
=
\hat H_+ 
- i \hat \Gamma
,
\ee
where
\be
 \hat H_\pm 
=
\frac{1}{2}
\left(
 \hat H \pm  \hat H^\dagger
\right)
=
\pm \hat H_\pm^\dagger 
,
\ee
and $\hat \Gamma = i \hat H_- = i
\left(
 \hat H -  \hat H^\dagger
\right)/2$ 
is a Hermitian operator, often referred as the \textit{decay rate operator}.
In our case, we obtain
\ba
 \hat H_+
 &=&
\hat\su_+ \hat{p}_x
+
\frac{1}{2}
  \left\{ 
	\huyy,
	\hat{p}_D^{-2}
  \right\}\!\hat{p}_x
 - \beta \hat{p}_D^{-2} \hat{p}_x 
,
 \label{Herm}\\
\hat \Gamma
&=&
i\, \hat\su_- \hat{p}_x
-
\frac{i}{2}
  \left[
	\huyy,
	\hat{p}_D^{-2}
  \right]\hat{p}_x
+
\mu_\text{dw} \hat I
\nn\\&=&
\hat \Gamma_0 + 
\frac{1}{2}
\frac{\dot \np}{\np} \hat I
,
 \label{Antiherm}
 \ea
where 
\ba
\hat \Gamma_0
&\dfn&
\hat \Gamma -
\frac{1}{2}
\frac{\dot \np}{\np} \hat I
\nn\\&=&
i\, \hat\su_- \hat{p}_x
-
\frac{i}{2}
  \left[
	\huyy,
	\hat{p}_D^{-2}
  \right]\hat{p}_x
+
\mu^{(0)}_\text{dw} \hat I
,
 \lb{AntihermO}\\
\hat\su_\pm 
&\dfn& 
\frac{1}{2} \left(\hat\su \pm \hat\su^\dag\right)
\nn\\&=&
\frac{1}{2} \left(
\hat{p}_D^{-2}
\hat{U}
\hat{p}_D^{2} \pm \hat{p}_D^{2}
\hat{U}
\hat{p}_D^{-2}
\right)
= \pm 
\hat\su_\pm^\dag
, \lb{e:sudef}
\ea
and
$\{~,~\}$ and $[~,~]$ are the anticommutator and commutator, respectively.
Using these, \eq \eqref{e:sudef} can be written in the form
\ba
\hat\su_+
&=&
\frac{1}{2}
\left(
\hat{p}_D^{-2}
\left\{
\hat{U},
\hat{p}_D^{2} 
\right\}
+
\hat{p}_D^{2}
\left[
\hat{U},
\hat{p}_D^{-2}
\right]
\right)
\nn\\&=&
\frac{1}{2}
\left(
\left\{
\hat{U},
\hat{p}_D^{2} 
\right\}
\hat{p}_D^{-2}
-
\left[
\hat{U},
\hat{p}_D^{-2}
\right]
\hat{p}_D^{2}
\right)
,\lb{e:sudef2p}\\
\hat\su_-
&=&
\frac{1}{2}
\left(
\hat{p}_D^{-2}
\left[
\hat{U},
\hat{p}_D^{2} 
\right]
-
\hat{p}_D^{2}
\left[
\hat{U},
\hat{p}_D^{-2}
\right]
\right)
\nn\\&=&
\frac{1}{2}
\left(
\left[
\hat{U},
\hat{p}_D^{2} 
\right]
\hat{p}_D^{-2}
-
\left[
\hat{U},
\hat{p}_D^{-2}
\right]
\hat{p}_D^{2}
\right)
, \lb{e:sudef2m}
\ea
which is more convenient for our calculations.

Let us turn our attention to the evolution equation for function $U$,
which must be also rewritten in a wave-mechanical form. 
From \eqs \eqref{eq:Ru6b} and \eqref{e:wtopsi} we obtain
\ba
\partial_t U
=
-
\mu_\text{zf}^{(0)} U
+
\np\,
\partial_y\! 
\left(
\overline{
\langle \textbf{x}| \hat{p}_y |\Psi \rangle 
\langle \Psi | \hat{p}_x |\textbf{x} \rangle 
}
\right)
,
\label{e:queq}
\ea
where an overscore denotes a zonal average per usual.

To conclude this section, 
we showed that drift waves 
can be described not only as waves but also as ``quanta'' (states in a Hilbert space)
which is somewhat similar to the de Broglie's wave-particle duality in quantum mechanics.
However,
one must emphasize that this analogy between fluctuation equations in turbulent flows
and a Schr\"odinger-type equation 
is not an exactly quantum-mechanical one.
The reason for this is that
the flow equation does not contain a Planck constant, but only
an analogue thereof, which is determined by pertinent scales of length, time and mass.
Up to a dimensionless coefficient, we can define
this effective Planck constant as
\be\lb{e:hbareff}
\hbar_\text{eff} \dfn
m_i \gyr^2 \omega_{ci} 
= 
\frac{m_i c_s^2}{\omega_{ci}} 
=
\frac{T_e}{\omega_{ci}}
,
\ee
where we used magnetized plasma's characteristic scales of length $\gyr$, time $\omega_{ci}^{-1}$,
and mass $m_i$, taken from Section \ref{s:moda}.
In the case of atmospheric fluids one could use, respectively, the planet's average radius, angular speed of planet's rotation,
and mass of a characteristic fluid parcel of planet's atmosphere.
For the sake of brevity,
the value of $\hbar_\text{eff}$
can be set to one, by rescaling and working in properly chosen units.

Nevertheless,
this is yet another example of macroscopic phenomena which have quantum-like features and  
can thus be described by virtue of notions and methods originating from quantum mechanics.
One of such formalisms is an analogue of quantum-statistical master equation approach in the theory 
of open quantum systems, which will be described in the next section.
While \eqs \eqref{eq:FlucQuantx} and \eqref{eq:FlucQuant} 
are simply a way of rewriting the original flow equations,
the next section's formalism is actually a generalization, which ushers in
a new physics.

\scn{Density operator formalism}{s:do}
In this section, we describe statistical approach based on the density operator, which is analogous to the von Neumann approach in quantum mechanics of mixed states, i.e., probabilistic mixtures of pure states \cite{sz13,sz14,sz14cor,ser15w,z15,sz15}.
In the matrix representation of a density operator, pure states are described by diagonal elements of a density matrix, whereas mixed states - by off-diagonal components thereof.
It is the latter which are regarded in quantum theory as an essentially quantum-mechanical effect, which does not usually have a classical counterpart.

When it comes to 
NH systems,
the density operator formalism is more general than the state-vector one, because
it allows the description of not only pure states, or their superpositions, but also mixed states.
This generalization is especially important, because dissipative effects are known to evolve some pure states
into mixed ones \cite{z15}.  
In other words, restricting ourselves to state vectors \textit{ab initio} would pose an implicit assumption of forbidding the transition
from pure to mixed states.
This assumption does not seem to be realistic  when it comes to open systems.

Besides,
if one were dealing with an equation for state vectors driven by a non-Hermitian Hamilton operator, such as \eq \eqref{eq:FlucQuant},
then one would arrive at complex eigenvalues of energy, which seem somewhat contradictory to the notion of energy \textit{per se}.  
On the contrary, in a density operator approach, no complex-valued energies occur:
the role of anti-Hermitian part of the Hamiltonian is to describe the time evolution of energy eigenvalues of its Hermitian part.

\sscn{Master equation: Non-sustainable evolution}{s:dome}
Let us regard the operator $|\Psi \rangle \langle  \Psi|$ as a special case of the
reduced
density operator $\nndo$.
From \eq \eqref{eq:FlucQuant} and its adjoint,
one can obtain an equation for $\nndo$,
which is thus called a master equation:
 \be
 i \partial_t \nndo 
=
 \hat H \nndo - \nndo\hat H^\dag 
 +
 i \disop
=
 [\hat H_+, \nndo] 
-
 i 
\left\{\hat \Gamma, \nndo\right\}
 + i \disop,
 \label{e:nndoevfull}
 \ee
 where 
\be\lb{e:disop}
\disop \to \np^{-1/2}
\left(
|\tilde \xi \rangle \langle \Psi|
 +
 |\Psi \rangle \langle \tilde \xi|
\right)
\ee
is a \textit{dissipator} operator, which is self-adjoint by construction.
Here, the arrow stands for ``for pure states tends to'',
which indicates that $\disop$ is a mixed-state generalization
of the operator $|\tilde \xi \rangle \langle \Psi|
 +
 |\Psi \rangle \langle \tilde \xi|$.
In that case,
one should use the hybrid approach,
which deals with systems which are driven by both Liouvillian (or Lindblad) and non-Hermitian Hamiltonian parts
of their master equations \cite{sz14}.

If
we neglect the dissipator term for the sake of simplicity,
we arrive at a canonical NH
master equation
\ba
\partial_t \nndo 
&=&
-i [\hat H_+, \nndo] 
- 
\left\{\hat\Gamma, \nndo\right\}
\nn\\&=&
-i [\hat H_+, \nndo] 
- 
\left\{\hat\Gamma_0, \nndo\right\}
-
\frac{\dot \np}{\np} \nndo
,
 \label{e:nndoevCut}
 \ea
where the NH operators are given by \eqs \eqref{Herm}-\eqref{AntihermO}.
Physical observables in this approach are defined as statistical averages: 
 \be
\av{\hat A}  \dfn \Tr (\hat A \nndo), 
 \lb{avdef}
 \ee
for a given observable's operator $\hat A$,
and similarly for correlation functions \cite{sz14cor}.

Let us recall \eq \eqref{e:queq}, which is supposed to supplement our master equation. 
Unlike the master equation, it is not an operator equation, therefore $U$ can be considered a c-number. 
Altogether we obtain
\ba
\partial_t \nnido 
&=&
-i [\hat H_+, \nnido] 
- 
\left\{\hat\Gamma_0, \nnido\right\}
,\\
\partial_t  U
&=&
-
\mu_\text{zf}^{(0)}  U
+
\partial_y 
\,
\overline{
\text{tr}\!\left( \hat{p}_y  \nnido \hat{p}_x 
\right)
}
, 
\ea
where we denoted
$\nnido
=
\np \nndo $.
In the last equation,
we introduced 
the partial trace operation `$\text{tr}$',
which is an algebraic matrix tracing or summation over all states of the type
$| \Psi_k \rangle \langle \Psi_k |$, but without averaging (integrating)
over  configuration space.

One can see that the $\np$-dependent term in \eq \eqref{Antiherm} can be
removed by  rescaling the  density operator.
Note that the averages are still computed with respect to $\nndo$,
therefore
$ 
\av{\hat A}  
= \np^{-1} \Tr (\hat A \nnido) = \np^{-1} \avi{\hat A} 
$, 
according to \eq \eqref{avdef}.
Recalling that the typical averages in our case, such as \eqref{e:ensdwb} and \eqref{e:endwb},
have a coefficient $\np$,
we can transform our density operator
$ 
\nndo \mapsto \nnido = \np \nndo
$, $
\av{\hat A}  
\mapsto
\avi{\hat A} = \np \av{\hat A}
$, 
then
drop the subscript `$\np$',
and assume 
$ 
\np \mapsto 1
$ 
from now on.
This contributes to the remarks made after \eq \eqref{e:mudw}:
the  $\np$-dependent term in the Hamiltonian disappears when working 
within the framework of a more general approach.

We thus finally obtain
the following set of evolution equations
\ba
\partial_t \nndo 
&=&
-i [\hat H_+, \nndo] 
- 
\left\{\hat\Gamma_0, \nndo\right\}
,
 \label{e:nndoev}\\
\partial_t  U
&=&
-
\mu_\text{zf}^{(0)}  U
+
\partial_y 
\,
\overline{
\text{tr}\!\left( \hat{p}_y  \nndo \hat{p}_x 
\right)
}
,
\label{e:udo} 
\ea
while the averages are defined according to \eq \eqref{avdef}.
For example,
using \eqs \eqref{e:ensdwb}-\eqref{e:enzfb},
we can define enstrophy and entropy as
the following statistical averages:  
\ba 
\mathcal{Z}_\text{dw} 
  &=&
	\frac{1}{2} 
\av{\hat{p}^{4}_D}
=
	\frac{1}{2} 
\Tr (\hat{p}^{4}_D \nndo)
 ,
 \label{e:ensdwbz}\\
  \mathcal{Z}_\text{zf}
  &=&
 \frac{1}{2}\int d y \, \uy^2
 ,\label{e:enszfbz}\\
  \mathcal{E}_\text{dw}
  &=&
	\frac{1}{2} 
\av{\hat{p}^{2}_D}	
=
	\frac{1}{2} 
\Tr (\hat{p}^{2}_D\, \nndo)
 ,\label{e:endwbz}\\
 \mathcal{E}_\text{zf}
 &=&
 \frac{1}{2}\int d y \, U^2 
 ,
 \label{e:enzfbz}
 \ea  
while the total values are given by \eqs \eqref{e:eetot}, per usual.

Finally, taking trace of \eq \eqref{e:nndoev}, 
one can obtain the following equation 
\be
\dot T_W
= - 2 \Tr(\hat \Gamma_0 \nndo) 
= - 2 \av{\hat\Gamma_0}
,
 \label{e:trev}
 \ee
supplemented with the initial condition
$
T_W (t=t_0) = 1
$,
where $ T_W (t) \dfn \Tr \, \nndo = \av{\hat I} $,
and
$t_0$ is the moment of time at which the environment switches on.
This equation clearly indicates that the trace of the operator 
$\nndo$ is not necessarily conserved during evolution.

This means that the drifton (sub)system experiences drain or loss of its degrees of freedom,
which can result in its total decay or critical instability.
While this  can indeed be the case in some systems,
it is not compulsory for all open systems.
In fact, next we are going to consider the possibility
of other type of evolution.

\sscn{Master equation: Sustainable evolution}{s:domes}

Dynamical systems, which can be described by non-Hermitian Hamiltonians, can follow two types
of evolution;
which can be referred as non-sustainable and sustainable, by analogy with some photobiological systems,
where this difference becomes striking \cite{z17adp}.
The former type is the one 
described by the density $\nndo$ obeying the master equation \eqref{e:nndoevCut}.
The sustainable type is the one described by the normalized density operator 
\be\lb{e:doans}
 \ndo = \nndo/ \Tr (\nndo) 
 , \quad 
 \ee
so that $\Tr (\ndo) = 1 $ at all times, which automatically removes the probability gain/loss problem \cite{sz13}.
This density would be a solution of the equation
 \ba
 \partial_t \ndo
 &=&
 -i [\hat H_+, \ndo]
 -
 \left\{\hat \Gamma_0, \ndo\right\}
 +
 2\, \Tr(\hat\Gamma_0 \ndo) \, \ndo
,
 \label{evolndo}
 \ea 
which is both nonlocal and nonlinear with respect to a density operator.
Fortunately, for practical computations, this equation can be transformed into a linear equation of
type \eqref{e:nndoev} by
using the ansatz \eqref{e:doans}.

For the sustainable type of evolution, 
physical observables are defined as
 \ba
\avo{\hat A} \dfn \Tr (\hat A \ndo)
,
 \lb{avavdef}
 \ea
for a given observable's operator $\hat A$,
and similarly for correlation functions \cite{sz14cor}.
Using the ansatz \eqref{e:doans}, one can relate the two types of averages:
 \be
\avo{\hat A} = \av{\hat A}/ \Tr (\nndo)
= \av{\hat A}/ \av{\hat I}
,
 \lb{e:avavo}
 \ee
which is often convenient for optimizing computations.

Furthermore,
recalling the mean-value origin of the last term
in \eq \eqref{eq:zRu6b}, the operator $\nndo$ must be replaced with $\ndo$ in the equation for the function $U$:
\be
\partial_t  U
=
-
\mu_\text{zf}^{(0)}  U
+
\partial_y 
\,
\overline{
\text{tr}\!\left( \hat{p}_y\, \ndo \hat{p}_x 
\right)
}
,
\label{e:undo}
\ee
according to remarks preceding \eq \eqref{avavdef}. 

Similarly to \eqs \eqref{e:ensdwbz}-\eqref{e:enzfbz}, 
one can define ``normalized'' enstrophy and entropy as
the following statistical averages:  
\ba 
\esds
  &=&
	\frac{1}{2} 
\avo{\hat{p}^{4}_D}
=
	\frac{1}{2} 
\Tr (\hat{p}^{4}_D \ndo)
 ,
 \label{e:ensdwbzs}\\
\eszs
  &=&
 \frac{1}{2}\int d y \, \uy^2
 ,\label{e:enszfbzs}\\
\ends
  &=&
	\frac{1}{2} 
\avo{\hat{p}^{2}_D}	
=
	\frac{1}{2} 
\Tr (\hat{p}^{2}_D\, \ndo)
 ,\label{e:endwbzs}\\
\enzs
 &=&
 \frac{1}{2}\int d y \, U^2 
 ,
 \label{e:enzfbzs}
 \ea  
where $U$ is now a solution of \eq \eqref{e:undo}.
While these definitions look very similar to their non-sustainable evolution 
analogues \eqref{e:ensdwbz}-\eqref{e:enzfbz},
one should notice the differences; such as the coefficients, functions of time,
which occur both in the definitions
themselves and in \eq \eqref{e:undo}.

To summarize Sections \ref{s:dome} and \ref{s:domes}: 
in the case of non-Hermitian Hamiltonian systems, one can have two types of evolution, each described by its own sets of equations.
In the case of non-sustainable evolution, this set consists of \eqs \eqref{e:nndoev} and \eqref{e:udo},
whereas in the case of sustainable evolution -- of \eqs \eqref{evolndo} and \eqref{e:undo}.
The choice between these types of NH evolution, or even the switch between them, is a process which is external to
the subsystem itself (it should also be remembered that we are dealing with reduced density operators here).
In other words, this choice must be considered a special case of the effect induced by the environment, which can also include the measuring
apparatus itself, as well as any pre- and post-selection measurement protocols.  
For example, in quantum optics, the calibrating and resetting of
fields inside optical fibers,
to prevent them from damage or to prepare for the next run, would be another example of sustainable-type evolution.

\sscn{Time evolution of averages}{s:doav}
Master equations are differential equations for operators, which makes  solving them a technically challenging task.
However, some physical information can be extracted without actually finding a density operator explicitly.
One of methods of such an extraction are equations for averages, which can be derived from a master equation.

Let us consider the non-sustainable type of NH evolution,
described by \eqs \eqref{e:nndoev} and \eqref{e:udo}. 
Multiplying \eq \eqref{e:nndoev} with various operators and tracing, one can obtain
 time evolution equations for corresponding averages 
as ordinary differential equations.
For example:
\ba
\partial_t \avz{\hat I}
&=& 
- 2 \avz{\hat\Gamma}
, \label{e:trevz}\\
\partial_t \avz{\hat{p}^{2}_D} 
&=&
-i \avz{
[
\hat{p}^{2}_D, \hat H_+
]
} 
- 
\avz{
\{\hat{p}^{2}_D, \hat\Gamma\}
}
\nn\\&=&
-i \avz{
[
\hat{p}^{2}_D, \hat H
]
} 
- 
2 \avz{
 \hat\Gamma \hat{p}^{2}_D
}
, \label{e:apd2}\\
\partial_t \avz{\hat{p}^{4}_D} 
&=&
-i \avz{
[
\hat{p}^{4}_D, \hat H_+
]
} 
- 
\avz{
\{\hat{p}^{4}_D, \hat\Gamma\}
}
\nn\\&=&
-i \avz{
[
\hat{p}^{4}_D, \hat H_+
]
} 
- 
\avz{
\hat{p}^{2}_D
[\hat{p}^{2}_D, \hat\Gamma]
}
\nn\\&&- 
\avz{
[\hat{p}^{2}_D, \hat\Gamma]
\hat{p}^{2}_D
}
, \label{e:apd4}\\
\partial_t \avz{\hat H_+} 
&=&
- 
\avz{
\{\hat\Gamma, \hat H_+\}
}
, \\
\partial_t \avz{\hat\Gamma} 
&=&
-i \avz{
[
\hat\Gamma, \hat H_+
]
} 
- 2
\avz{
\hat\Gamma 
}
,
 \ea
where we used identities mentioned after \eq \eqref{e:hamtot}
and the identity 
$\{ \hat A \hat B, \hat C \} = \hat A [\hat B, \hat C] + [\hat A, \hat C] \hat B$.
If some of the resulting differential equations form a nontrivial closed subset,
they can then
be solved for the corresponding averages as functions of time.
Otherwise, one has to apply approximations or iteration methods to the full set.


 \scn{Phase-space formulation: Non-sustainable evolution}{s:ww}
%
Master equations of the types described in the previous section are differential equations for operators.
This creates a formidable technical problem when it comes to  solving them.
As a result, there are a number of methods which allow us to solve them, 
each with a different degree of accuracy or completeness.
Solving the equations for averages is one of the approaches - it allows us
to compute averages
as solutions of a set of differential equations for functions, instead of
dealing with operators.
However, in the case of an infinite-dimensional Hilbert space, the number of such equations is not
necessarily finite or reducible to something simple,
but dependent on other symmetries of the problem.
The second method, which is to be discussed in this section,  
is based on mapping between
functions in the quantum phase space formulation and Hilbert space operators in the Schr\"odinger representation 
\cite{wbook,w32,gr46,hos84}, which was adapted for density operators for various NH systems \cite{cv07,gs11,bc13,bcl15},
including zonal-flow models specifically \cite{wo09,p16,rd16}. 

There also exists a subtle point when it comes to the completeness of the phase space description of density operators for NH systems including zonal-flow ones.
As mentioned at the beginning of \Sec~\ref{s:do},
it is important to preserve and extract, as fully as possible, 
an essentially non-classical piece of information 
about time evolution of a probabilistic mixture of states of NH-driven systems \cite{z15}. 
This is straightforward within the framework of the Hilbert space matrix formulation of a density operator 
from \Sec~\ref{s:do}, where the structure is clear: this information is encoded mostly in the off-diagonal elements of a density matrix, as we know it from quantum mechanics of mixed states.
It is far less obvious how one can separate ``off-diagonal'' effects from ``diagonal'' ones in the phase space formulation of a density operator, in presence of additional 
approximations which are technically inevitable to obtain definite results. 

\sscn{Wigner-Weyl transform}{s:wwmap}
Let us introduce  
the Weyl symbol of the density operator $\nndo$ as the integral \cite{gr46}:
\ba
\mathfrak W (\nndo)
&=&
W(\textbf{x}, \textbf{p},t) 
\nn\\&\dfn&
\int d^2 s \Exp{-i \mathbf{p}
\cdot 
\mathbf{s}} \langle\mathbf{x}+\mathbf{s/2}| \nndo|\mathbf{x}-\mathbf{s/2}\rangle
,
~~~~~~
\label{DensM}
\ea
from which the following property  
\be
W(\textbf{x}, \textbf{p},t)=W(\textbf{x}, -\textbf{p},t)
\label{DensME}
\ee
can be easily deduced in the case of real-valued wavefunctions.

Correspondingly,
Weyl symbols for operators in the previous section will turn into functions (c-numbers).
For example,
using the resolution of identity,
we obtain for the
operator $\hat p_x$:
\bw
\ba
 \mathfrak W (\hat p_x)
 &=&
 \int d ^2 s \Exp{-i \mathbf{p}\cdot \mathbf{s}} \langle\mathbf{x}+\mathbf{s/2}| \hat p_x|\mathbf{x}-\mathbf{s/2}\rangle
= 
  \int d^2 s \, d^2 p' \Exp{-i \mathbf{p}\cdot \mathbf{s}} \langle\mathbf{x}+\mathbf{s/2}| \hat p_x|\textbf{p}'\rangle\langle \textbf{p}'|\mathbf{x}-\mathbf{s/2}\rangle
\nn\\&=&
\frac{1}{2 \pi}\int d^2 p' p'_x \int d^2 s \exp[i \textbf{s}\cdot (\textbf{p}'-\textbf{p})]
  =
  \int d^2 p'p'_x \delta(\textbf{p}'-\textbf{p})
  = p_x
,
\label{invpx}
 \ea
for the operator $\hat p_D^{-2} $: 
 \ba
 \mathfrak W(\hat p_D^{-2})
&=&
  \int d^2 s \, d^2 p' \Exp{-i \mathbf{p}\cdot \mathbf{s}} \langle\mathbf{x}+\mathbf{s/2}| \frac{1}{\hat p^2 +1}|\textbf{p}'\rangle\langle \textbf{p}'|\mathbf{x}-\mathbf{s/2}\rangle
\nn\\&=&
\frac{1}{2 \pi}\int \frac{d^2 p'}{ p'^2 +1} \int d^2 s \exp[i \textbf{s}\cdot (\textbf{p}'-\textbf{p})]
= 
  \int \frac{d^2 p'}{ p'^2 +1} \delta(\textbf{p}'-\textbf{p})
  = p_D^{-2}
	,
 \label{invpd}
 \ea 
\ew
for the similarity transform operator and its combinations,
using \eqs \eqref{e:sudef}-\eqref{e:sudef2m}:
\ba
\mathfrak W(\hat U)
&=& U 
,\\ 
\mathfrak W(\huyy)
&=&
\partial^2_y \, \mathfrak W(\hat U)
= \uyy
,\lb{invu}\\
 \mathfrak W(\hat\su)
&=&
p_D^{-2} \star  \mathfrak W(\hat U) \star p_D^{2}
\dfn
\su
, \\
 \mathfrak W(\hat\su_+)
&=&
\frac{1}{2}
\left(
p_D^{-2} \star  \mathfrak W(\hat U) \star p_D^{2}
+
p_D^{2} \star  \mathfrak W(\hat U) \star p_D^{-2}
\right)
\nn\\&=&
\frac{1}{2}
\left(
\mobs{U, p_D^{2}} \star p_D^{-2}
-
i
\mob{U, p_D^{-2}} \star p_D^{2}
\right)
\nn\\&\dfn&
\su_+
,\lb{e:sudef3p}\\
 \mathfrak W(\hat\su_-)
&=&
\frac{1}{2}
\left(
p_D^{-2} \star  \mathfrak W(\hat U) \star p_D^{2}
-
p_D^{2} \star  \mathfrak W(\hat U) \star p_D^{-2}
\right)
\nn\\&=&
\frac{i}{2}
\left(
\mob{U, p_D^{2}} \star p_D^{-2}
-
\mob{U, p_D^{-2}} \star p_D^{2}
\right)
\nn\\&\dfn&
i \su_-
,\lb{e:sudef3m}
\ea
where 
$\mob{,}$ and $\mobs{,}$ are, respectively, sine and cosine Moyal brackets  defined in Appendix B. 
 
Furthermore, using the Moyal product rule \eqref{MoyR} 
we obtain for operator products, such as
$ 
 \mathfrak W(\hat{p}_x\hat{p}_D^{-2}) = p_x \star p_D^{-2}=p_x(1+i \hat{\mathfrak L}/2)p_D^{-2}
 =
p_x/p_D^2
$, 
where $p_D^2 = p_x^2+p_y^2+1$.
Similarly $\mathfrak{W}(\hat{U}\hat{p}_x)=  U p_x$. 
Other properties of the Moyal product, which will be used in what follows,
such as the associativity \eqref{AssM},
are listed in Appendix B. 
 
Performing the Weyl transform of \eq \eqref{e:nndoev},
we obtain the equation
 \be 
 \partial_t W
 =
\mob{H_+,W} - 
\mobs{\Gamma_0,W} 
,
 \label{PseuDE}
 \ee
where
\ba
 H_+
 &\dfn&
\mathfrak W(\hat H_+)
\nn\\&=&
\su_+ p_x
+
\frac{1}{2}
\mobs{\uyy, 1/p_D^2} p_x
 - \beta p_x /p_D^2 
,
 \label{HermW}\\
\Gamma_0
 &\dfn&
\mathfrak W(\hat \Gamma_0)
\nn\\&=&
-
\su_- p_x
+
\frac{1}{2}
\mob{\uyy, 1/p_D^2} p_x
+
\mu_\text{dw}^{(0)}
,
 \label{AntiHW}
 \ea
according to \eqs \eqref{Herm}, \eqref{Antiherm}, \eqref{invpx}-\eqref{e:sudef3m}.

Furthermore, let us consider an equation for function $U$, which supplements the master equation \eqref{PseuDE}.
Since \eq \eqref{e:udo} is not an operator equation, it is only the last term which should be rewritten in terms of Wigner function.
We obtain
\ba
\partial_t  U
+
\mu_\text{zf}^{(0)}  U
&=&
\partial_y 
\int \frac{d^2 p}{(2 \pi)^2} p_y \star \zfl{W} \star p_x
\nn\\&=&
\partial_y 
\int \frac{d^2 p}{(2 \pi)^2} p_x p_y \zfl{W} 
,
\label{e:udoww}
\ea
where  we took into account in the last step
that
the zonal-averaged Wigner function
$\zfl W$ does not depend on $x$ 
and satisfies the condition \eqref{DensME}.

Finally,
the integrals \eqref{e:ensdwbz}-\eqref{e:enzfbz}
are transformed into the form
\ba 
\mathcal{Z}_\text{dw} 
  &=&
	\frac{1}{2} 
\av{\hat{p}^{4}_D}
=
\frac{1}{2} 
\int \frac{d^2p}{(2\pi)^2} 
d^2 x
\, p^{4}_D \star {W}
 ,
 \label{e:ensdwbz2}\\
  \mathcal{Z}_\text{zf}
  &=&
 \frac{1}{2}\int d y \, \uy^2
 ,\label{e:enszfbz2}\\
  \mathcal{E}_\text{dw}
  &=&
	\frac{1}{2} 
\av{\hat{p}^{2}_D}	
=
\frac{1}{2} 
\int \frac{d^2p}{(2\pi)^2} 
d^2 x
\, p^{2}_D \star {W}
 ,\label{e:endwbz2}\\
 \mathcal{E}_\text{zf}
 &=&
 \frac{1}{2}\int d y \, U^2 
 ,
 \label{e:enzfbz2}
 \ea 
and the total values are as defined in \eq \eqref{e:eetot}.


\sscn{Eikonal approximation}{s:eiko}
Equation \eqref{PseuDE} remains
difficult to solve, except in a few cases when the Moyal products series can be truncated.
In our case, it seems that no truncation is possible, therefore one has to resort to the eikonal or geometrical-optics approximation \eqref{e:eiko}, which is analogous to a leading-order WKB approximation
with respect to the effective Planck constant \eqref{e:hbareff}.
Then
\eqs \eqref{e:sudef3p} and \eqref{e:sudef3m} become
approximately
\ba
\su_+
&=&
U
, \\ 
\su_-
&=&
\frac{1}{2}
\left(
\pob{U, p_D^{2}} p_D^{-2}
-
\pob{U, p_D^{-2}} p_D^{2}
\right)
\nn\\&=& 
2 p_y U_1/p_D^2
,\lb{e:sudef4m}
\ea
and 
the whole system \eqref{PseuDE}-\eqref{e:udoww} simplifies to
a system of integro-differential equations
 \ba
&&
 \partial_t W = \pob{\mathcal{H},W}
- 2 \mathcal{G} W
- 2 \mu_\text{dw}^{(0)} W
,
 \label{LowO}
\\&&
\partial_t  U
+
\mu_\text{zf}^{(0)}  U
=
\partial_y 
\int \frac{d^2 p}{(2 \pi)^2} p_x p_y \zfl{W} 
,
\label{e:udowww}
\ea
where  
\ba
\mathcal H 
&=&
p_x U
+
p_x \uyy/p_D^2
- \beta p_x /p_D^2 
,
 \label{HermP}\\
\mathcal G
&=&
- 2 p_x p_y U_1/p_D^2
+
\frac{1}{2}
\pob{\uyy, p_x/p_D^2}
\nn\\&=&
- 2 p_x p_y U_1/p_D^2
-
p_x p_y U_3/p_D^4
,
 \label{AntihP}
\ea
where $\pob{,}$ is a canonical Poisson bracket as defined in Appendix B. 

Notice that in the formula for $\mathcal H$, all derivatives of $U$ are even, while 
for
$\mathcal G$ - all are odd, which is similar to even-odd function classification in quantum mechanics.
In this case, this is caused
by a discrete $\mathbb Z_2$ symmetry with respect to the mirror transformation 
$x \to x,\, y \to -y$,
which leaves invariant both $\mathcal H$ and
$\mathcal G$. 
This symmetry thus supplements the parity-time reversal symmetry
$\vc x \to - \vc x, \, t \to -t$, of
the system \eqref{LowO}-\eqref{AntihP}, 
which occurs when $\mu_\text{dw}^{(0)} = \mu_\text{zf}^{(0)} = 0$.

By zonal averaging of \eqs \eqref{LowO} and \eqref{e:udowww}, we obtain the following
equations
\ba
&&
\partial_t \zfl W 
= \pob{\mathcal{H},\zfl W} - 2 \mathcal G \zfl W
- 2 \mu_\text{dw}^{(0)} \zfl W
,
 \label{LowO2}
\\&&
\partial_t  U
=
-
\mu_\text{zf}^{(0)}  U +
\partial_y\!
\int \frac{d^2 p}{(2 \pi)^2} p_x p_y \zfl{W} 
,
\label{e:udoww2}
\ea
where we also used \eqs \eqref{e:mudw} and \eqref{e:mu0}.
The former of these equations belongs to a class of generalized WKE models, some other examples to be
found in \cite{rd16}.

Finally,
in the eikonal approximation,
the integrals \eqref{e:ensdwbz2}-\eqref{e:enzfbz2}
can be simplified to
\ba 
\mathcal{Z}_\text{dw} 
  &=&
\frac{1}{2} 
\int \frac{d^2p}{(2\pi)^2} 
d^2 x
\, p^{4}_D  {W}
= 
\frac{1}{2} 
\int \frac{d^2p}{(2\pi)^2} 
d y
\, p^{4}_D  \zfl{W}
 ,~~~~~~~
 \label{e:ensdwbze}\\
  \mathcal{Z}_\text{zf}
  &=&
 \frac{1}{2}\int d y \, \uy^2
 ,\label{e:enszfbze}\\
  \mathcal{E}_\text{dw}
  &=&
\frac{1}{2} 
\int \frac{d^2p}{(2\pi)^2} 
d^2 x
\, p^{2}_D {W}
= 
\frac{1}{2} 
\int \frac{d^2p}{(2\pi)^2} 
d y
\, p^{2}_D \zfl{W}
 ,\label{e:endwbze}\\
 \mathcal{E}_\text{zf}
 &=&
 \frac{1}{2}\int d y \, U^2 
 ,
 \label{e:enzfbze}
 \ea 
while the total values are as defined in \eq \eqref{e:eetot}, as before.
The rates of 
these values are 
\ba
\Der{}{t} 
\mathcal{Z}_\text{dw} 
  &=&
\frac{1}{2} 
\int \frac{d^2p}{(2\pi)^2} 
d y
\, p^{4}_D  
\, 
\partial_t \zfl{W}
\nn\\&=&
- 2 \mu^{(0)}_\text{dw} \mathcal{Z}_\text{dw}
- 
\int \frac{d^2p}{(2\pi)^2} 
d y
\, p_x p_y U_3 \zfl W
 ,~~
 \label{e:rensdwbze}\\
\Der{}{t} 
  \mathcal{Z}_\text{zf}
  &=&
\int d y \, \uy \,\partial_y \partial_t U
=
- \int d y \, \uyy \, \partial_t U
\nn\\&=&
- 2 \mu^{(0)}_\text{zf} \mathcal{Z}_\text{zf}
+
\int \frac{d^2p}{(2\pi)^2} 
d y
\, p_x p_y U_3 \zfl W
 ,\label{e:renszfbze}\\
\Der{}{t} 
  \mathcal{E}_\text{dw}
  &=&
\frac{1}{2} 
\int \frac{d^2p}{(2\pi)^2} 
d y
\, p^{2}_D  
\, \partial_t \zfl{W}
\nn\\&=&
- 2 \mu^{(0)}_\text{dw} \mathcal{E}_\text{dw}
+
\int \frac{d^2p}{(2\pi)^2} 
d y
\, p_x p_y \uy \zfl W
 ,\label{e:rendwbze}\\
\Der{}{t} 
 \mathcal{E}_\text{zf}
 &=&
\int d y \, U \,\partial_t U 
\nn\\&=&
- 2 \mu^{(0)}_\text{zf} \mathcal{E}_\text{zf}
-
\int \frac{d^2p}{(2\pi)^2} 
d y
\, p_x p_y \uy \zfl W
 ,
 \label{e:renzfbze}
 \ea 
where we used evolution equations \eqref{LowO2}, \eqref{e:udoww2},
and formulae from Appendix B. 

Using these expressions for rates,
one can immediately see that  total enstrophy and total energy \eqref{e:eetot}
evolve in our model according to the formulae
\ba
\Der{}{t} 
\mathcal{Z}_\text{tot} 
 &\dfn&
\Der{}{t}
\mathcal{Z}_\text{dw}
+
\Der{}{t}
\mathcal{Z}_\text{zf}
\nn\\&=&
-2
\left(
\mu^{(0)}_\text{dw} \mathcal{Z}_\text{dw}
+
\mu^{(0)}_\text{zf} \mathcal{Z}_\text{zf}
\right)
,\\
\Der{}{t} 
\mathcal{E}_\text{tot} 
 &\dfn&
\Der{}{t}
\mathcal{E}_\text{dw}
+
\Der{}{t}
\mathcal{E}_\text{zf}
\nn\\&=&
-2
\left(
\mu^{(0)}_\text{dw} \mathcal{E}_\text{dw}
+
\mu^{(0)}_\text{zf} \mathcal{E}_\text{zf}
\right)
,
\ea
therefore, 
depending on values of $\mu$'s, our system's enstrophy and energy can evolve in different ways.
Let us consider the following special cases:

\begin{itemize} 
\item	\textit{Overall conservation}
occurs when 
\be\lb{e:mu0}
\mu_\text{dw}^{(0)} = \mu_\text{zf}^{(0)} = 0
.
\ee   
In this case, 
both energy and enstrophy can flow from the drift-wave component to the zonal-flow one,
and back,
but do not leave the system:
\ba
\mathcal{Z}_\text{tot} (t) &=& \mathcal{Z}_\text{tot} (0) = \text{const}
,\\
\mathcal{E}_\text{tot} (t) &=& \mathcal{E}_\text{tot} (0) = \text{const}
.
\ea

\item	\textit{Overall exponential gain or loss}
occurs when 
\be\lb{e:mumu}
\mu_\text{dw}^{(0)} = \mu_\text{zf}^{(0)} = \lambda/2
,
\ee
where $\lambda$ is a real-valued constant.   
In this case, 
the
total
energy and enstrophy of our system both increase (decrease) if $\lambda$ is negative (positive),  
at an exponential rate:
\ba
\mathcal{Z}_\text{tot} (t) &=& \mathcal{Z}_\text{tot} (0) \exp{(- \lambda t)}
,\\
\mathcal{E}_\text{tot} (t) &=& \mathcal{E}_\text{tot} (0) \exp{(- \lambda t)}
,
\ea
which ultimately leads to either critical instability 
or to the
complete depletion of the system. 
Thus, our system has a finite lifetime $\tau = 1/\lambda$ in this case. 

\end{itemize}

To conclude, an eikonal approximation reveals a number of important features of dissipative processes
induced by drift-wave turbulence,
including the behaviour of average values of energy and vorticity of its drifton and zonal-flow components.

\scn{Phase-space formulation: Sustainable evolution}{s:wws}
In this section, we adopt a Wigner-Weyl formalism in the case of the normalized density operator $\ndo$.
In the
computations that follow, an ansatz \eqref{e:doans} will be of great assistance,
because it offers us a shortcut 
for our calculations,
then use the results related to non-sustainable evolution.  

\sscn{Wigner-Weyl transform}{s:wwmaps}
Let us introduce  
the Weyl symbol of the normalized density operator $\ndo$ as the ratio \cite{pz18}:
\be
\mathfrak W (\ndo)
\dfn
\rho (\textbf{x}, \textbf{p},t)  = 
\mathfrak W (\nndo) / T_W (t)
,
\ee
where
\be
T_W (t) =
\int 
\frac{d^2 p}{(2 \pi)^2} 
d^2 x \,
\mathfrak W (\nndo)
,
\ee
and 
$\mathfrak W (\nndo)$ is given by \eq \eqref{DensM}.
Therefore,
an ansatz \eqref{e:doans} can be applied in the form 
\ba
\rho (\textbf{x}, \textbf{p},t)  
&=& 
\ssw W (\textbf{x}, \textbf{p},t) / T_{\ssw W} (t)
, \lb{e:doansW}\\
T_{\ssw W} (t)
&=&
\int 
\frac{d^2 p}{(2 \pi)^2} 
d^2 x \, {\ssw W} (\textbf{x}, \textbf{p},t)
, \lb{e:trevsW}
\ea
where $\ssw W(\textbf{x}, \textbf{p},t)$
is a solution of
the equation
 \be 
 \partial_t \ssw W
 =
\mob{\sss H_+, \ssw W} - 
\mobs{\sss \Gamma_0, \ssw W} 
,
 \label{PseuDEs}
 \ee
where
\ba
 \sss H_+
 &\dfn&
H_+ \bigr|_{U = \sss U}
\nn\\&=&
\sss\su_+ p_x
+
\frac{1}{2}
\mobs{\sss U_2, 1/p_D^2} p_x
 - \beta p_x /p_D^2 
,~~
 \label{HermWs}\\
\sss\Gamma_0
 &\dfn&
\Gamma_0 \bigr|_{U = \sss U}
\nn\\&=&
-
\sss\su_- p_x
+
\frac{1}{2}
\mob{\sss U_2, 1/p_D^2} p_x
+
\mu^{(0)}_\text{dw}
,
 \label{AntiHWs}
 \ea
where
\ba
\sss\su_+
&=&
\frac{1}{2}
\left(
\mobs{\sss U, p_D^{2}} \star p_D^{-2}
-
i
\mob{\sss U, p_D^{-2}} \star p_D^{2}
\right)\!,~~~~~~~\lb{e:sudef3ps}\\
\sss\su_-
&=&
\frac{1}{2}
\left(
\mob{\sss U, p_D^{2}} \star p_D^{-2}
-
\mob{\sss U, p_D^{-2}} \star p_D^{2}
\right)\!,\lb{e:sudef3ms}
\ea
where function $\sss U$,
according to \eqs \eqref{e:undo} and \eqref{e:doansW}, is a solution of the equation
\ba
\partial_t  \sss U
+
\mu_\text{zf}^{(0)}  \sss U
= 
\frac{1}{T_{\ssw W}}
\partial_y \int \frac{d^2 p}{(2 \pi)^2}\, p_y \star \ssw{\zfl{W}} \star p_x
,
\label{e:udowws}
\ea
where $\ssw{\zfl{W}} $ is a zonal average of the solution of \eq \eqref{PseuDEs}.

Furthermore,  
averages \eqref{e:ensdwbzs}-\eqref{e:enzfbzs}
take the form
\ba 
\esds
  &=&
\frac{1}{2 T_{\ssw W}}
\int \frac{d^2p}{(2\pi)^2} 
d^2 x
\, p^{4}_D \star \ssw W
 ,
 \label{e:ensdwbz2s}\\
\eszs
  &=&
 \frac{1}{2}\int d y \, \sss U_1^2
 ,\label{e:enszfbz2s}\\
\ends
  &=&
\frac{1}{2 T_{\ssw W}}
\int \frac{d^2p}{(2\pi)^2} 
d^2 x
\, p^{2}_D \star \ssw{W}
 ,\label{e:endwbz2s}\\
\enzs
 &=&
 \frac{1}{2}\int d y \, \sss U^2 
 ,
 \label{e:enzfbz2s}
 \ea 
where we used the ansatz \eqref{e:doansW}.

\sscn{Eikonal approximation}{s:eikos}
By analogy with Section \ref{s:eiko}, we obtain that
\eqs \eqref{e:sudef3ps} and \eqref{e:sudef3ms} become
in a leading-order approximation
\ba
\sss\su_+
&=&
\sss U
, \\ 
\sss\su_-
&=&
\frac{1}{2}
\left(
\pob{\sss U, p_D^{2}} p_D^{-2}
-
\pob{\sss U, p_D^{-2}} p_D^{2}
\right)
\nn\\&=& 
2 p_y \sss U_1/p_D^2
,\lb{e:sudef4ms}
\ea
therefore,
the system \eqref{PseuDEs}-\eqref{e:udowws} simplifies to
a system of integro-differential equations
 \ba
&&
 \partial_t \ssw W = \pob{\sss{\mathcal{H}}, \ssw W}
- 2 \sss{\mathcal{G}} \ssw W
- 2 \mu^{(0)}_\text{dw} \ssw W
,
 \label{LowOs}
\\&&
\partial_t  \sss U
+
\mu_\text{zf}^{(0)}  \sss U
=
\frac{1}{T_{\ssw W}}
\partial_y 
\int \frac{d^2 p}{(2 \pi)^2} p_x p_y \ssw{\zfl{W}} 
,
\label{e:udowwws}
\ea
where  
\ba
\sss{\mathcal H} 
&=&
p_x \sss U
+
p_x \sss U_2/p_D^2
- \beta p_x /p_D^2 
,
 \label{HermPs}\\
\sss{\mathcal G}
&=&
- 2 p_x p_y \sss U_1/p_D^2
+
\frac{1}{2}
\pob{\sss U_2, p_x/p_D^2}
\nn\\&=&
- 2 p_x p_y \sss U_1/p_D^2
-
p_x p_y \sss U_3/p_D^4
.
 \label{AntihPs}
\ea
Therefore, 
performing zonal averaging of \eqs \eqref{LowOs}, we obtain the zonal-averaged
master
equation
\ba
\partial_t \ssw{\zfl{W}}  
&=& \pob{\sss{\mathcal{H}}, \ssw{\zfl{W}} } - 2 \sss{\mathcal G} \ssw{\zfl{W}} 
- 2 \mu^{(0)}_\text{dw} \ssw{\zfl{W}}
,
 \label{LowO2s}
\ea
while \eq \eqref{e:udowwws} stands as is.

It is useful to compute the trace of the auxiliary density operator in this approximation.
Using \eqs \eqref{e:trev}, \eqref{e:doansW}, \eqref{e:trevsW}, \eqref{AntihPs}
and \eqref{LowO2s},
we obtain
\be\lb{e:trevsW2}
\dot T_{\ssw W} 
=
- 2 \mu^{(0)}_\text{dw} T_{\ssw W} 
-
2 {\mathcal G}_{W}
,
\ee
where
\ba
{\mathcal G}_{W} (t)
&\dfn&
\int \frac{d^2p}{(2\pi)^2} 
d y
\, \sss{\mathcal G}  \sss{\zfl{W}}
\nn\\&=&
-
\int \frac{d^2p}{(2\pi)^2} 
d y
\, 
\frac{p_x p_y}{p_D^2}
\left(
2 U_1 +
\frac{U_3}{p_D^2}
\right) \sss{\zfl{W}}
,~~~
\lb{e:truncgamo}
\ea
is an average of the truncated decay rate operator,
$\hat \Gamma_0
-
\mu^{(0)}_\text{dw} \hat I$,
with respect to the auxiliary (non-normalized) density operator.

Finally,
in the eikonal approximation,
averages \eqref{e:ensdwbz2s}-\eqref{e:enzfbz2s}
can be simplified to
\ba
\esds 
  &=&
\frac{1}{2 T_{\ssw W}}
\int \frac{d^2p}{(2\pi)^2} 
d y
\, p^{4}_D  \sss{\zfl{W}}
 ,
 \label{e:ensdwbzes}\\
\eszs
  &=&
 \frac{1}{2}\int d y \, \sss U_1^2
 ,\label{e:enszfbzes}\\
\ends
  &=&
\frac{1}{2 T_{\ssw W}}
\int \frac{d^2p}{(2\pi)^2} 
d y
\, p^{2}_D \sss{\zfl{W}}
 ,\label{e:endwbzes}\\
\enzs
 &=&
 \frac{1}{2}\int d y \, \sss U^2 
 ,
 \label{e:enzfbzes}
 \ea 
therefore, rates of 
these values can be computed,
using the evolution equations \eqref{e:udowwws}, \eqref{LowO2s}, 
 \eqref{e:trevsW2},  
and formulae from Appendix B, 
as
\ba
\Der{}{t} 
\esds
&=&
\frac{1}{2 }
\int \frac{d^2p}{(2\pi)^2} 
d y
\, p^{4}_D  
\left[
\Der{}{t} 
\left(
\frac{
1
}{T_{\ssw W}}
\right)
\sss{\zfl{W}}
+
\frac{
1
}{T_{\ssw W}}
\partial_t
\sss{\zfl{W}}
\right]
\nn\\&=&
2
{\mathcal G}_{\rho}
\esds
-
\frac{1}{T_{\ssw W}}
\int \frac{d^2p}{(2\pi)^2} 
d y
\, p_x p_y \sss U_3 \ssw{\zfl{W}}
 ,
 \label{e:rensdwbzes}\\
\Der{}{t} 
\eszs
  &=&
\int d y \, \sss U_1 \,\partial_y \partial_t \sss U
=
- \int d y \, \sss U_2 \, \partial_t \sss U
\nn\\&=&
- 2 \mu^{(0)}_\text{zf} \eszs
+
\frac{1}{T_{\ssw W}}
\int \frac{d^2p}{(2\pi)^2} 
d y
\, p_x p_y \sss U_3 \ssw{\zfl{W}} 
 ,\label{e:renszfbzes}\\
\Der{}{t} 
\ends
&=&
\frac{1}{2 }
\int \frac{d^2p}{(2\pi)^2} 
d y
\, p^{2}_D  
\left[
\Der{}{t} 
\left(
\frac{
1
}{T_{\ssw W}}
\right)
\sss{\zfl{W}}
+
\frac{
1
}{T_{\ssw W}}
\partial_t
\sss{\zfl{W}}
\right]
\nn\\&=&
2
{\mathcal G}_{\rho}
\ends
+
\frac{1}{T_{\ssw W}}
\int \frac{d^2p}{(2\pi)^2} 
d y
\, p_x p_y \sss U_1 \ssw{\zfl{W}} 
 ,\label{e:rendwbzes}\\
\Der{}{t} 
\enzs
 &=&
\int d y \, \sss U \,\partial_t \sss U 
\nn\\&=&
- 2 \mu^{(0)}_\text{zf} \enzs
-
\frac{
1}{T_{\ssw W}}
\int \frac{d^2p}{(2\pi)^2} 
d y
\, p_x p_y \sss U_1 \ssw{\zfl{W}} 
 ,
 \label{e:renzfbzes}
 \ea 
where 
we used
${\mathcal G}_{\rho} (t) \dfn
{\mathcal G}_{W} / T_{\ssw W}$
to denote
an average of the truncated decay rate operator
with respect to the main (normalized) density operator.

Using these formulae,
one can see that ``normalized''
total enstrophy and total energy \eqref{e:eetot}
evolve in our model according to the formulae
\ba
\Der{}{t} 
\ssss{\mathcal{Z}}_\text{tot} 
 &\dfn&
\Der{}{t}
\esds
+
\Der{}{t}
\eszs
\nn\\&=&
-2
\left(
{\mathcal G}_{\rho}  \esds
+
\mu^{(0)}_\text{zf} \eszs
\right)
,\\
\Der{}{t} 
\ssss{\mathcal{E}}_\text{tot} 
 &\dfn&
\Der{}{t}
\ends
+
\Der{}{t}
\enzs
\nn\\&=&
-2
\left(
{\mathcal G}_{\rho} \ends
+
\mu^{(0)}_\text{zf} \enzs
\right)
,
\ea
where we used the total values defined by \eqs \eqref{e:eetot} as analogy, per usual.
Notice that, unlike their analogues from Section \ref{s:eiko},
these rates do not depend on $\mu^{(0)}_\text{dw}$.

\scn{Conclusion}{s:con}
In this paper,
we have presented a statistical mechanical approach to describing dissipative phenomena in zonal flows
of plasmas and atmospheric fluids, such as drift waves and Rossby waves,
which is based on Landau-von Neumann's density operator in a theory of open quantum systems.
This became possible due to an occurrence  of Hilbert space associated with an
electric potential
or stream function, which is regarded as the fundamental Hilbert space of the theory.
This results in a formal mapping between flow and wave equations,
which allows us to describe zonal flows and associated phenomena as macroscopic wave-mechanical effects.
As a consequence of this mapping,
flow equations can be rewritten as Schr\"odinger-like equations, with two important differences:
for dimensionality purposes,
one uses an effective Planck constant whose value is not necessarily equal to the quantum-mechanical Planck constant,
and the resulting Hamiltonian operator is not necessarily Hermitian.

The second feature requires us to treat the entire theory 
from within the framework of the non-Hermitian Hamiltonian approach,
where anti-Hermitian parts of Hamiltonian operators usually
describe an effect of the environment.
Fortunately,  such an approach has been already developed and applied to various open systems.
According to this formalism, one has to generalize from state vectors to density operators, 
because pure states do not necessarily stay pure during the time evolution in the presence of dissipation
and noise. 
Moreover,
the
density operator approach prevents the occurrence of complex-valued energies.
In this approach, energies are always real-valued, and refer to the subsystem itself,
whereas any anti-Hermitian components describe the effects of
the environment upon this subsystem (decay rates, \textit{et cetera}). 

Thus, after deriving a Hamiltonian operator and evolution equations for state vectors
in our DW/RW models,
we made a transition from state vectors to density operators,
introduced NH master equations and defined observables;
including the enstrophy and energy of both the waves and zonal flow.
The upshot is that
two types
of density operator's evolution exist,
these can be referred to as non-sustainable and sustainable, by analogy with some photobiological systems
where they can be visualized.

The non-sustainable type is the one described by the non-normalized density operator. 
During such evolution, an open (sub)system experiences drain or loss of its degrees of freedom,
which can result in its total decay or critical instability.
While this can indeed be the case in some systems,
it is not a compulsory feature of all open systems.

The sustainable type is the one described by the density operator
which is normalized at all-times.
This automatically removes the problem of probability's gain or loss in NH systems,
thus enhancing their stability.
An example of such stability would be an environment-assisted stability in photobiological systems.
Consequently, various observables, including those related to enstrophy and energy,
behave in a way different from the non-sustainable case.

Because  we are dealing with reduced density operators;
a selection of one or another type of NH evolution, or even the switch between them at some point in time, is a 
process external
to the subsystem itself.
In fact, this process is neither unitary nor continuous.
It can be considered a special case of the effect induced by the environment, 
which can also include the measuring
apparatus itself.  

Furthermore, to establish a method for solving master equations for a given DW/RW model,
we considered a phase-space formulation of the theory.
We introduced relevant Weyl-Wigner transforms
and rewrote evolution equations and observables in the Moyal form.
We also studied a leading-order approximation of the Wigner approach,
for both types of evolution,
which is analogous to the eikonal or geometrical approximation in optics or WKB approximation in quantum mechanics.

As it turns out,
the statistical-mechanical density operator formalism
shows itself 
to be an approach to dissipative phenomena related to zonal flows,
which has clear foundations and notions emanating from quantum mechanics.
It also allows us
to take into account the wave-mechanical effects in the above-mentioned systems,
thus producing more realistic descriptions thereof. 
As such, the formalism can also be extended 
to other dissipative systems in plasma and atmospheric physics,
which allow the wave-mechanical analogy and non-Hermitian Hamiltonian operator description.\\

\noindent
\textbf{Abbreviations}\\ 
The following abbreviations are used in this manuscript:\\

\noindent 
\begin{tabular}{@{}ll}
DW & Drift wave, drifton\\
HME & Hasegawa-Mima equation\\
NH & Non-Hermitian Hamiltonian\\
RW & Rossby wave\\
WKB & Wentzel-Kramers-Brillouin\\
WKE & Wave kinetic equation\\
ZF & Zonal flow
\end{tabular}

\begin{acknowledgments}This research is supported by Department of Higher Education and Training of South Africa and in part by National Research Foundation of South Africa.
Proofreading of the manuscript by P. Stannard is greatly appreciated.
\end{acknowledgments}

\appendix

\scn{DERIVATION OF VORTICITY}{s:appw}

In the case of effectively two-dimensional systems,
one can define vorticity as a projection of
the vorticity pseudovector on the third axis.
We therefore assume
\be
w \dfn
\mathbf w 
\cdot \ez
=
(\vena \times \mathbf {v_{\perp}})
\cdot \ez
,
\label{DerVort}
\ee
where 
$ \mathbf w 
=
\vena \times \mathbf {v_{\perp}}
$
is the vorticity pseudovector,
and
$\mathbf {v_{\perp}}$
is the
ion's velocity in the $(x,y)$ plane.
Using \eq \eqref{e:vappr},
we obtain
\ba\lb{e:vorve}
\mathbf w
=
\mathbf w^{(1)} + 
\mathbf w^{(2)} +
\mathbf w^{(3)}
,
\ea
where
\ba
\mathbf w^{(1)}
&=&
-\frac{1}{B} \vena \times (\vena \phi \times \ez)
,\\
\mathbf w^{(2)}
&=&
-
\frac{1}{\omega_{ci}B}\vena \times \partial_t \vena \phi
= 0
,\\
\mathbf w^{(3)}
&=&
\frac{1}{\omega_{ci}B^2}\vena \times 
\left\{
[(\vena \phi \times \ez)\cdot \vena ] \vena \phi
\right\}
.
\ea
In the components' notations,
we can write these expressions as
\ba
(\mathbf w^{(1)})_k
&=&
- \frac{1}{B}\epsilon_{mkl} \epsilon_{mj3} \partial_l \partial_j \phi
 =
 \frac{1}{B}\delta_{k3} \partial_m \partial_m \phi
,\\
(\mathbf w^{(3)})_k
&=&
\frac{1}{\omega_{ci}B^2}
\epsilon_{kmn} \epsilon_{ij3}
\partial_i \partial_n \phi \; \partial_m \partial_j 
\phi
\\ \nn 
&=&
\frac{1}{\omega_{ci}B^2}
\delta_{k 3}(\partial _m \partial _n \phi \; \partial _m \partial _n \phi
-
\partial _n \partial _n \phi \; \partial _m \partial _m \phi
)
,
\label{thirdVort}
\ea
where
we used the following properties of the Levi-Civita symbol in three dimensions:
\ba 
\epsilon_{ijk} \epsilon_{imn}
&=&
\delta_{jm}\delta _{kn} 
-
\delta _{jn} \delta _{km}
, \label{e:levic1}\\
\epsilon_{ijk} \epsilon_{lmn}
&=&
\delta_{il}(\delta _{jm} \delta _{kn}
-
\delta _{jn} \delta _{km})
\nn\\&&
-
\delta_{im}(\delta _{jl} \delta _{kn}
-
\delta _{jn} \delta _{kl})
\nn\\&&
+
\delta_{in}(\delta _{jl} \delta _{km}
-
\delta _{jm} \delta _{kl}
).
\label{e:levic2}
\ea  

{\bf\small Appendix B: WIGNER-WEYL FORMALISM}\\

The Weyl symbol $A(\mathbf{x},\mathbf{p})$ for any given operator $\hat A$ is defined as
\bw
\ba
 A(\mathbf{x},\mathbf{p})
&=&
 \int d ^n s \Exp{-i \mathbf{p}\cdot \mathbf{s}} \langle\mathbf{x}+\mathbf{s/2}| \hat A|\mathbf{x}-\mathbf{s/2}\rangle.
 \label{WeylS}
\\
 \mathcal{A}(\textbf{x},\textbf{x}^\prime)
 &=&
 \frac{1}{(2\pi)^n} 
\int d^np \exp{[-i\textbf{p} \cdot (\textbf{x}^\prime - \textbf{x})]} \,
A \left( \frac{\textbf{x}^\prime + \textbf{x}}{2}, \textbf{p}\right),
 \label{MEinCR}
 \ea
\ew
 in particular
 \be 
 \mathcal{A}(\textbf{x},\textbf{x})
 =
 \int \frac{d^n p}{(2 \pi)^n }A(\textbf{x},\textbf{p}).
 \label{MEinP}
 \ee
Completeness condition
 \be
 \int_{- \infty}^{ \infty}d^n p|\mathbf{p} \rangle \langle\mathbf{p}|=1.
 \label{Compl}
 \ee
 Delta function
 \be
 \delta(\textbf{x})=\frac{1}{2 \pi}\int_{- \infty}^{ \infty}d^n p \Exp{i \textbf{p}\cdot \textbf{x}}.
 \label{delta}
 \ee
Inner product of two different variables
 \be
 \langle \mathbf{x},\mathbf{p} \rangle
 =
 \frac{1}{\sqrt{2 \pi}}\Exp{i \textbf{p}\cdot \textbf{x}}.
 \label{}
 \ee
 Moyal product rule:
 for any $\hat C= \hat A \hat B$, the corresponding Weyl symbols satisfy
 \be
 C(\textbf{x},\textbf{p})=A(\textbf{x},\textbf{p}) \star B(\textbf{x},\textbf{p}),
 \label{MoyR}
 \ee
 where $\star$ is the Moyal product
defined as
\be
 A(\textbf{x},\textbf{p}) \star B(\textbf{x},\textbf{p})
 \dfn
  A(\textbf{x},\textbf{p}) \Exp{i \hat{ \mathfrak{L}}/2} B(\textbf{x},\textbf{p}),
 \label{MoyP}
 \ee
where 
$ 
 \hat{ \mathfrak{L}}
 \dfn
 \overleftarrow{\partial_{\textbf{x}}}\cdot  \overrightarrow{\partial_{\textbf{p}}}
 -
 \overleftarrow{\partial_{\textbf{p}}}\cdot  \overrightarrow{\partial_{\textbf{x}}}
$ 
is the Janus operator,
and 
\be
A \hat{ \mathfrak{L}} B
\dfn
\pob{A,B}
=
 \partial_{\textbf x}A \, \partial_{\textbf p}B
 -
 \partial_{\textbf p}A \, \partial_{\textbf x}B
\label{CanP} \ee 
is the canonical Poisson bracket.

The Moyal product is associative
 \be
 A\star B\star C
 \dfn
( A\star B)\star C
 =
 A\star (B\star C)
,
 \label{AssM}
 \ee
and becomes an ordinary product inside the phase space integrals
 \be
 \int d^nx d^np A \star B
 =
 \int d^nx d^np AB
,
 \label{MPregP}
 \ee
provided integrands vanish at integration boundaries.
Using the Moyal product,
it is convenient to define the Moyal or sine bracket
 \be
 \mob{A,B}
 \dfn 
 -i(A \star B-B \star A)
 =
 2A \sin(\hat{ \mathfrak{L}}/2)B
,
 \label{SineB}
 \ee 
which is the Wigner map of the commutator,
\be
\left[\hat A, \hat B \right]
\mapsto
i \mob{\mathfrak W (\hat A), \mathfrak W (\hat B)} =
i \mob{A,B}
,
\ee
as well as the symmetric Moyal or Groenewold-Baker's cosine bracket 
\be
\mobs{A,B}
 \dfn 
 A \star B+B \star A
 =
 2A \cos(\hat{ \mathfrak{L}}/2)B
,
 \label{CosB}
\ee
which is the Wigner map of the anti-commutator
\be
\left\{\hat A, \hat B \right\}
\mapsto
\mobs{\mathfrak W (\hat A), \mathfrak W (\hat B)} =
\mobs{A,B}
,
\ee
where $\mathfrak W (\hat A)$ denotes the Weyl transform of $\hat A$.

When assuming an eikonal or leading-order WKB approximation,
these brackets have the following properties
\ba
\mob{A,B}
&\mapsto&
 A\hat{ \mathfrak{L}}B= \pob{A,B}
+
{\cal O} (\hbar_\text{eff}^2)
,\nn\\
\mobs{A,B}
&\mapsto&
 2AB
+
{\cal O} (\hbar_\text{eff}^2)
,
\label{e:eiko}
\ea
where the effective Planck constant is defined in \eq \eqref{e:hbareff}.
 
Finally, let us give some useful formulae for our case.
If configuration space is two-dimensional and phase-space functions $A$ and $B$ do not depend on a coordinate $x$,
we obtain
\ba
\pob{A,B} &=& \partial_y A \, \partial_{p_y} B - \partial_{p_y} A \, \partial_y B
,\\
f 
\pob{A,B} &=& - B \, \partial_{p_y} f 
\, \partial_y A 
\nn\\&&+ 
{\cal D} (\partial_{p_y}, \partial_{y})
,
\ 
\forall \, f = f (\vc p),
\ea 
where 
$ 
{\cal D} (\partial)
$ 
are total derivative terms with respect to arguments listed in braces.
Such terms can be omitted when working
inside reduced phase space integrals: $\int d^2 p\, dy \,{\cal D} (\partial) = 0$,
assuming that physical values vanish at phase space borders.
For instance,
for the functions
\ba
\mathcal H 
&=&
p_x U
+
p_x \uyy/p_D^2
- \beta p_x /p_D^2 
,
 \nn\\
\mathcal G
&=&
- 2 p_x p_y U_1/p_D^2
-
p_x p_y U_3/p_D^4
,
\nn
\ea
we obtain the following
identity:
\ba
&&\int \frac{d^2p}{(2\pi)^2} 
d y
\, p^{2 n}_D  
\left(
\pob{\mathcal{H},\zfl W} - 2 \mathcal G \zfl W
\right)
\nn\\&&\qquad
=
2
\int \frac{d^2p}{(2\pi)^2} 
d y
\, 
p_x p_y 
p^{2 (n-1)}_D 
\nn\\&&\qquad\quad\times
\left[
(1-n)
\left(
U_1
+
\frac{U_3 }{p^{2}_D}
\right)
+
U_1
\right]
\zfl W
,~~~~~~
\ea
where $n$ being integer,
hence
\ba
&&
\int \frac{d^2p}{(2\pi)^2} 
d y
\, p^{4}_D  
\left(
\pob{\mathcal{H},\zfl W} - 2 \mathcal G \zfl W
\right)
\nn\\&&\qquad\qquad\qquad
=
- 2
\int \frac{d^2p}{(2\pi)^2} 
d y
\, p_x p_y U_3 \zfl W
,~~~\\&&
\int \frac{d^2p}{(2\pi)^2} 
d y
\, p^{2}_D  
\left(
\pob{\mathcal{H},\zfl W} - 2 \mathcal G \zfl W
\right)
\nn\\&&\qquad\qquad\qquad
=
2
\int \frac{d^2p}{(2\pi)^2} 
d y
\, p_x p_y U_1 \zfl W
,
\ea
because neither $U$ nor $\zfl W$ depend on $x$.

\end{document}